\documentclass[%
 reprint,
 amsmath,amssymb,
 aps,
longbibliography
]{revtex4-2}

\AtBeginDocument{
 \newwrite\bibnotes
 \def\bibnotesext{Notes.bib}
 \immediate\openout\bibnotes=\jobname\bibnotesext
 \immediate\write\bibnotes{@CONTROL{REVTEX41Control}}
 \immediate\write\bibnotes{@CONTROL{
apsrev41Control,author="08",editor="1",pages="1",title="0",year="1"}}
 \if@filesw
 \immediate\write\@auxout{\string\citation{apsrev41Control}}%
 \fi
}

\usepackage[utf8]{inputenc}
\usepackage{graphicx}
\usepackage{xcolor}
\usepackage[colorlinks=true,citecolor=blue]{hyperref}
\usepackage[caption=false]{subfig}
\usepackage{color}
\usepackage{cancel}
\usepackage{tabularx}
\usepackage{diagbox}
\usepackage{soul}
\usepackage{comment}
\newtheorem{proposition}{Proposition}

\begin{document}
\title{Information-theoretic analysis of temporal dependence in discrete stochastic processes: Application to precipitation predictability}
\author{Juan De Gregorio}
\email[]{juan@ifisc.uib-csic.es}
\author{David S\'anchez}
\author{Ra\'ul Toral}
\affiliation{
Institute for Cross-Disciplinary Physics and Complex Systems IFISC (UIB-CSIC), Campus Universitat de les Illes Balears, E-07122 Palma de Mallorca, Spain
}
\date{\today}
\begin{abstract}

Understanding the temporal dependence of precipitation is key to improving weather predictability and developing efficient stochastic rainfall models. We introduce an information-theoretic approach to quantify memory effects in discrete stochastic processes and apply it to daily precipitation records across the contiguous United States. The method is based on the predictability gain, a quantity derived from block entropy that measures the additional information provided by higher-order temporal dependencies. This statistic, combined with a bootstrap-based hypothesis testing and Fisher's method, enables a robust memory estimator from finite data. Tests with generated sequences show that this estimator outperforms other model-selection criteria such as Akaike Information Criterion and Bayesian Information Criterion. Applied to precipitation data, the analysis reveals that daily rainfall occurrence is well described by low-order Markov chains, exhibiting regional and seasonal variations, with stronger correlations in winter along the West Coast and in summer in the Southeast, consistent with known climatological patterns. Overall, our findings establish a framework for building parsimonious stochastic descriptions, useful when addressing spatial heterogeneity in the memory structure of precipitation dynamics, and support further advances in real-time, data-driven forecasting schemes.

\end{abstract}
\maketitle

\textbf{Predicting precipitation and other environmental variables depends critically on understanding how future conditions are determined from past events. Here, we present an information-theoretic framework that quantifies temporal dependencies directly from data and discuss an application to daily precipitation records across the contiguous United States. The approach provides a statistical robust way to detect memory effects in finite sequences, revealing regional and seasonal differences in rainfall correlations. By identifying when and where simple stochastic descriptions remain reliable, this framework supports responsible forecasting practices that balance accuracy, efficiency, and interpretability. These principles could extend to a broader class of complex systems where short-term predictability is essential.}

\section{Introduction}
\label{sec:intro}

Stochastic processes provide a concise, probabilistic framework for describing how complex systems evolve over time without the need to model every microscopic interaction. By focusing on the probabilities of transitioning between states~\cite{cox17,Doob}, it is possible to capture essential randomness and simplify the representation of otherwise intricate dynamics, thus being able to make statistical predictions about the possible outcomes of the process.

In the simplest memoryless processes, transition rates to future states depend neither on past outcomes nor on the current state, so that no historical information can improve prediction. A typical example of this kind of processes is a coin toss. However, many real‐world systems exhibit temporal correlations that violate this assumption: their future behavior depends on preceding events~\cite{han82}. To accommodate such dependencies, one may adopt an $m$th‐order (or ``higher‐order'') Markov representation, where the probability of the next state depends only on the results of $m$ previous realizations of the process, rather than its entire history~\cite{raf85}. Here, we will refer to $m$ as the process's ``memory'', noting that $m=1$ recovers the traditional Markov property~\cite{markov} and $m=0$ corresponds to independent, identically distributed (iid) variables.

In many cases, the inclusion of memory provided by the higher-order Markov approach offers a better statistical representation of a process, improving predictability. This approach has been widely applied in a large variety of fields, including, but not limited to, weather forecast~\cite{wil99}, linguistics~\cite{PhysRevLett.74.4559}, human navigation on the web~\cite{sin14} and DNA sequence analysis~\cite{narlikar}.

To fix ideas, let us consider a random variable $X$ with a finite number $L$ of possible outcomes, $\beta_0,\ldots,\beta_{L-1}$ and probability distribution 
\begin{equation}
P(X)=\lbrace p(\beta_i),\quad i=0,\ldots,L-1 \rbrace,  
\label{eq:prob_dist}
\end{equation}
where $p(\beta_i)$ is the probability that $X$ takes the value $\beta_i$. 
By repeating realizations of $X$ over time, we generate a stochastic process $\lbrace X_t \rbrace$, where $t=0,1,2,\ldots$ denotes the time index of the repetition in appropriate units. Furthermore, we restrict our analysis to stationary processes.

To analyze the evolution of the process, it is essential to examine how past outcomes influence the likelihood of future events. In particular, the $u$th-order transition probability $p(\beta_{i_{u+1}}|\beta_{i_{1}},\ldots,\beta_{i_{u}})$ quantifies the likelihood of observing the state $\beta_{i_{u+1}}$, 
given that the process was in the sequence of states $\beta_{i_{1}},\ldots,\beta_{i_{u}}$ over the previous $u$ time steps.
We say that a process has $u$th-order correlations if
\begin{equation}
p(\beta_{i_{u+1}}|\beta_{i_{1}}, \beta_{i_{2}},\ldots,\beta_{i_{u}}) \neq p(\beta_{i_{u+1}}|\beta_{i_{2}},\ldots,\beta_{i_{u}}),
\end{equation}
for some $i_1,\ldots,i_{u+1}$.

A stochastic process $\lbrace X_t \rbrace$ is said to have order or memory $m\geq 1$ if the transition probabilities of order $u\geq m$ satisfy
\begin{equation}
p(\beta_{i_{u+1}}|\beta_{i_{1}},\ldots,\beta_{i_{u}})=p(\beta_{i_{u+1}}|\beta_{i_{u-(m-1)}},\ldots,\beta_{i_{u}}).
\label{eq:high_markov}
\end{equation}
This is equivalent to stating that, for such systems, transition probabilities of order $u\geq m$ do not offer additional information beyond what is already captured by the $m$th-order transitions.

Incorporating a larger number of past states into the modeling of a process increases computational complexity, as it introduces a greater number of independent parameters. This added complexity can reduce estimation accuracy, especially when transition probabilities are not known a priori and must be inferred from finite data. Therefore, modeling a process as a higher-order Markov chain requires a balance between the accuracy of parameter estimation and the amount of information that past states can provide. It is therefore essential to determine the intrinsic memory of the process as the minimal value of $m$ that satisfies Eq.~\eqref{eq:high_markov}.

Common approaches for estimating the memory of a process include model selection criteria such as the Akaike Information Criterion (AIC)~\cite{aka98} and the Bayesian Information Criterion (BIC)~\cite{kat81}. However, these methods have notable limitations. AIC is known to favor overly complex models, whereas BIC tends to favor overly simplistic ones~\cite{wea99}. Additionally, both approaches are inherently model-dependent: they are designed to select the best model from a predefined set based on specific optimization criteria~\cite{ton75}. However, this selection process does not guarantee that the chosen model faithfully captures the true dynamics of the underlying process.

In this paper, we propose an information-theoretic approach designed to provide a deeper understanding of the process dynamics by explicitly analyzing temporal dependencies within the system. This method not only yields a more accurate estimate of the process memory, but also enables a step-by-step assessment of how past outcomes influence future evolution.
We demonstrate the practical applicability of the proposed approach for analyzing memory effects in stochastic processes from finite sequences with simulated data.

We apply this methodology to analyze real-world records of daily precipitation occurrence in the contiguous United States.
Precipitation is an especially interesting case because it exhibits strong seasonal and regional variability~\cite{w17172529}, and the persistence of daily conditions determines the length of wet and dry spells with direct societal and ecological impacts~\cite{https://doi.org/10.1002/eco.2518,lesk2016influence}. 
Although the proposed framework provides only a statistical forecast based on observed temporal dependencies, rather than a physically based prediction, it offers a complementary, data-driven perspective on short-term predictability. In particular, by quantifying memory and correlations across seasons and regions, our approach can guide model development by identifying situations where simpler, lower-parameter models are sufficient for specific locations and months, which can help reduce computational costs for physical simulations~\cite{reder2025estimating}.

The paper is organized as follows. In Section~\ref{sec:pred_gain}, we introduce the concept of predictability gain and establish its main properties, linking it to block entropy and conditional mutual information. Section~\ref{sec:memory_est} presents the proposed memory-estimation methodology based on hypothesis testing and bootstrap resampling, and evaluates its performance against AIC and BIC using simulated data. In Section~\ref{sec:prec_seq}, we apply the method to daily precipitation sequences across the contiguous United States, analyzing spatial and seasonal variability in estimated memory and temporal correlations. Finally, Section~\ref{sec:conclusions} summarizes the main conclusions and discusses potential applications to other domains involving complex stochastic dynamics. Appendixes~\ref{sec:appA}–\ref{app:C} contain the mathematical proofs of the propositions presented in the main text, additional theoretical results, and further details on the numerical implementation and data analysis.

\section{Predictability gain}
\label{sec:pred_gain}

Central to the proposed memory estimation method is the concept of block entropy, a generalization of Shannon entropy to sequences of consecutive outcomes of the discrete random variable $X$. Specifically, the block entropy of size $r\geq 1$ reads,
\begin{equation}
 H_r = -\sum_{i_1,\ldots,i_r=0}^{L-1}p(\beta_{i_1},\ldots,\beta_{i_r})\ln(p(\beta_{i_1},\ldots,\beta_{i_r})),
 \label{eq:entropy}
\end{equation}
with the convention $H_0 \equiv 0$.

An important connection arises from the relationship between memory and block entropy.
It has been proven that a stochastic process has memory $m\geq 0$ if and only if $H_r$ is a linear function of $r$ for $r\geq m$~\cite{juan}.
This result implies that the (negative) second discrete derivative of the block entropy, given by
\begin{equation}
\mathcal{G}_u = -(H_{u+2}-2H_{u+1}+H_u),
\label{eq:pred_gain}
\end{equation}
for integer $u\geq 0$, vanishes for all $u\geq m$.

\begin{figure}[h]
 \centering

 \begin{minipage}{\linewidth}
 \begin{minipage}{0.05\linewidth}
  \raggedright \textbf{(a)}
 \end{minipage}
 \begin{minipage}{0.9\linewidth}
  \includegraphics[width=\linewidth]{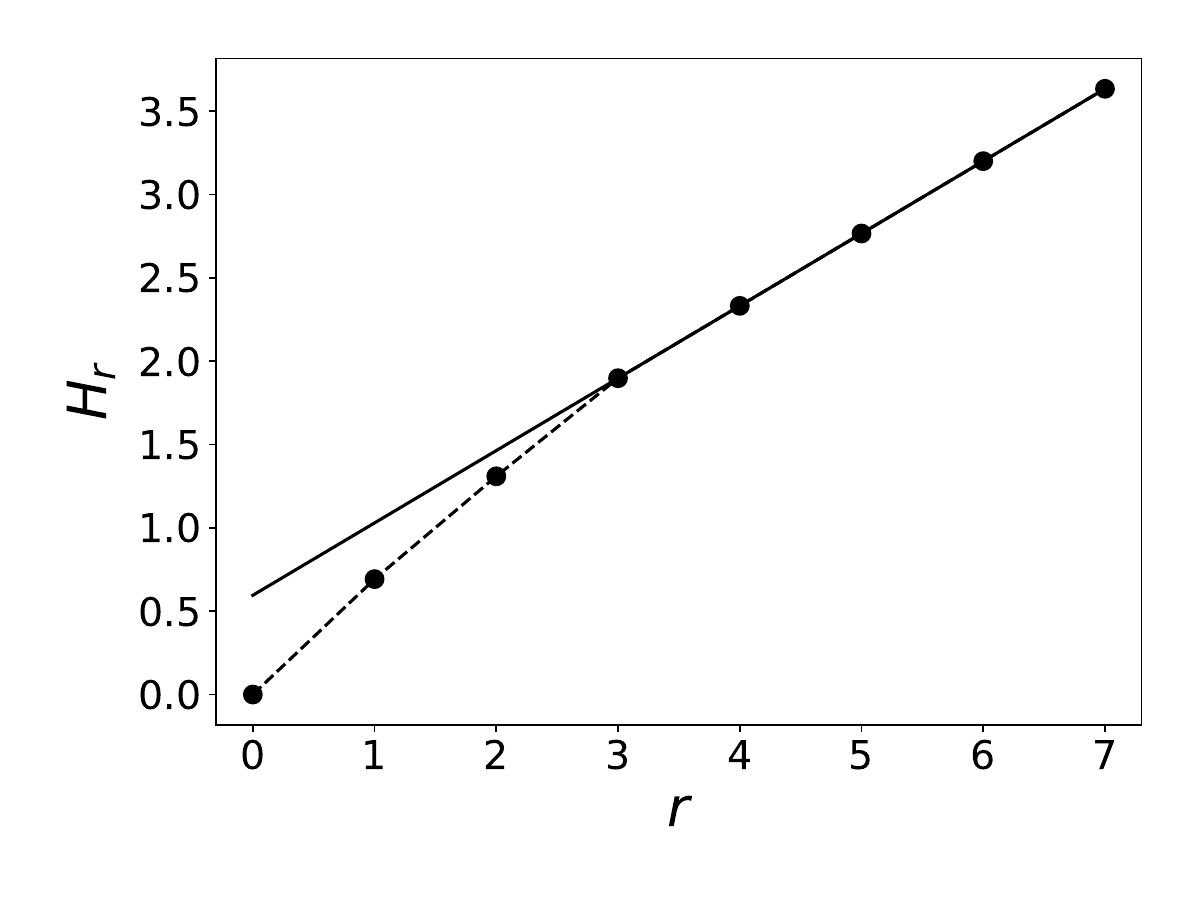}
 \end{minipage}
 \end{minipage}

 \vspace{2ex}

 \begin{minipage}{\linewidth}
 \begin{minipage}{0.05\linewidth}
  \raggedright \textbf{(b)}
 \end{minipage}
 \begin{minipage}{0.9\linewidth}
  \includegraphics[width=\linewidth]{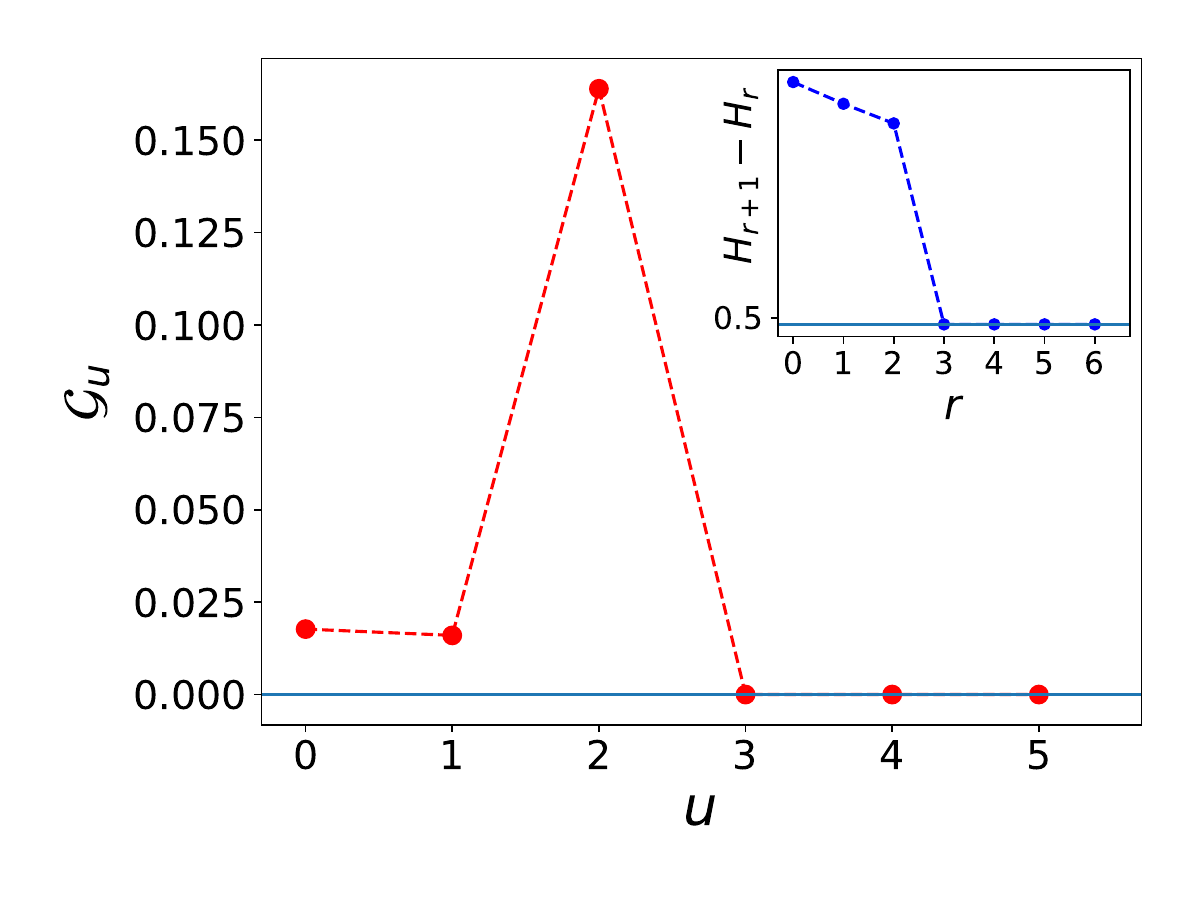}
 \end{minipage}
 \end{minipage}

\caption{Block entropy (a) and predictability gain (b) for a binary system with possible states $(\beta_0=0,\beta_1=1)$ and memory $m=3$. The inset in panel (b) displays the first discrete derivative of the block entropy. In order to avoid any spurious bias, the set of $8$ transition probabilities $p(0|\beta_{i_1},\beta_{i_2},\beta_{i_3})$ for $i_1,i_2,i_3=0,1$ has been chosen randomly from a uniform distribution. The complementary probabilities are then set as $p(1|\beta_{i_1},\beta_{i_2},\beta_{i_3})=1-p(0|\beta_{i_1},\beta_{i_2},\beta_{i_3})$. Once these probabilities have been chosen, known analytical expressions have been used to compute the block entropies.
\label{fig:example}}
\end{figure}

Therefore, the memory $m$ can be equivalently defined as
\begin{equation}
m = \text{min}(\lbrace \eta: \mathcal{G}_u = 0,\text{ for all } u\geq\eta \rbrace).
\label{eq:m}
\end{equation}
As an example, we depict in Fig.~\ref{fig:example} the block entropy and predictability gain for a
binary ($L=2$) system with memory $m=3$, based on one set of transition probabilities
drawn randomly from a uniform distribution.
As predicted, panel (a) confirms that the block entropy is linear for $r\geq 3$. This is further supported by panel (b), where $\mathcal{G}_u=0$ for $u\geq 3$. Additionally, the inset in panel (b) shows that the first discrete derivative of the block entropy, $H_{r+1}-H_r$,
decreases until $r=3$, after which it remains constant.

It should be noted that, for a different set of transition probabilities, the curves in Fig.~\ref{fig:example} would change quantitatively; however, the same conclusions discussed above regarding the memory and the behavior of the block entropy and its derivatives would still hold.

It can be shown that $\mathcal{G}_u$ is equivalent to a conditional mutual information~\cite{cover, crutchfield2}, and can therefore be expressed as
 \begin{equation}
 \begin{split}
 \label{eq:pred_gain2}
  \mathcal{G}_u = &\sum_{i_0,\ldots,i_{u}=0}^{L-1} p(\beta_{i_0},\ldots,\beta_{i_{u}}) \times \\
  &D_{\text{KL}}(P(X|\beta_{i_0},\beta_{i_1},\ldots,\beta_{i_{u}})||P(X|\beta_{i_1},\ldots,\beta_{i_{u}})),
  \end{split}
 \end{equation}
where $D_{\text{KL}}$ is the Kullback-Leibler divergence between conditional probabilities:
\begin{equation}
\label{eq:kl}
 D_{\text{KL}}(P(X|y)||P(X|z)) = \sum_{i=0}^{L-1} p(\beta_i|y)\ln \left(\dfrac{p(\beta_i|y)}{p(\beta_i|z)}\right).
\end{equation}
 
The expression given by Eq.~\eqref{eq:pred_gain2} reveals that $\mathcal{G}_u$ quantifies the average amount of information gained by considering $(u+1)$th-order transition probabilities instead of those of order $u$, for $u\geq 0$. For this reason, it is referred to as predictability gain~\cite{crutchfield2}, and satisfies the properties presented in Section~\ref{sec:properties} and proven in Appendix~\ref{sec:appA}, which indicate that this measure not only provides for a method to determine the memory value of a system, as suggested by Eq.~\eqref{eq:m}, but also allows for a precise quantification of the temporal correlations within a process.

\subsection{Properties}
\label{sec:properties}

\begin{proposition}
\label{prop:additive}
The predictability gain is additive: the amount of information gained
when considering $k$th-order transition probabilities instead of $u$th-order transitions, 
\begin{equation}
\begin{split}
G(u\to k)&\equiv \sum_{i_1,\ldots,i_{k}=0}^{L-1}p(\beta_{i_1},\ldots,\beta_{i_{k}}) \times \\ 
&D_{\text{KL}}(P(X|\beta_{i_1},\ldots,\beta_{i_{k}})||P(X|\beta_{i_{k-u+1}},\ldots,\beta_{i_{k}})),
\label{eq:cummulative_gain}
\end{split}
\end{equation}
for $k>u$, can be calculated as
\begin{equation}
 G(u\rightarrow k) = \sum_{l=u}^{k-1}\mathcal{G}_l.
\end{equation}
\end{proposition}

Proposition~\ref{prop:additive} demonstrates that the predictability gained from $u$th-order to $k$th-order transitions can be computed sequentially by accumulating the information gained at each intermediate step: from order $u$ to $u+1$, then from $u+1$ to $u+2$, and so on, up to order $k$.

\begin{proposition}
\label{prop:total_pred_gain}
The total predictability gain defined as
\begin{equation}
G_T \equiv \sum_{l=0}^{\infty} \mathcal{G}_l,
\label{eq:total_gain}
\end{equation}
can be computed as
\begin{equation}
G_T = H_1-h,
\label{eq:G_T}
\end{equation}
with $h=\lim_{r\to\infty}\dfrac{H_r}{r}$ the entropy rate of the system.
\end{proposition}

Given that the entropy rate quantifies the uncertainty in the next state of the process conditioned on its entire past, Proposition~\ref{prop:total_pred_gain} shows that the total predictability gain corresponds to the reduction in uncertainty obtained by moving from a memoryless description that ignores temporal correlations, to a representation that fully incorporates the history of the process.
An application of Eq.~\eqref{eq:G_T} can be found in Ref.~\cite{montemurro2011}, where it is used to quantify the information contributed by word ordering in different languages.

While the present work focuses on the analysis of temporal correlations within a single process, related information-theoretic approaches have recently addressed analogous issues in multivariate settings by introducing the partial information rate decomposition, a framework for characterizing dynamically shared information between a target process and multiple source processes~\cite{nrwj-n8lj}.

The predictability gain is sometimes referred to as active information storage and quantifies the amount of information from the system's past that is actively used to predict its next state~\cite{LIZIER201239,mediano2021extendedtaxonomyinformationdynamics}. It has been widely applied in neuroscience and the analysis of complex dynamical systems~\cite{BARA2023105315,ANTONACCI2025129437}, just to mention a few representative examples. 

\begin{proposition}
\label{prop:bounded}
The predictability gain is bounded:
\begin{equation}
0 \leq \mathcal{G}_u \leq \ln(L), \quad u\geq 0.
\label{eq:bound_pred}
\end{equation}
\end{proposition}

\begin{proposition}
\label{prop:pred_gain_distance}

For a process with memory $m\geq 1$, $\mathcal{G}_{m-1}$ measures the Euclidean distance in the $(r,H_r)$ plane between $H_r$ and $\mathcal{H}(r)$ at $r=m-1$, where $\mathcal{H}(r)$ is the straight line that fulfills
\begin{equation}
\mathcal{H}(r) = H_r, \quad r\geq m.
\end{equation}
\end{proposition}

Proposition~\ref{prop:pred_gain_distance} implies that $\mathcal{G}_{m-1}$ can be seen as a measure of how close a process with memory $m$ is to one with memory $m-1$. This makes $\mathcal{G}_{m-1}$ a useful criterion for deciding whether a lower-order approximation is justified. If its value is sufficiently small, the process may be effectively described with reduced memory, allowing for a simpler representation and lower computational cost without significantly compromising accuracy.

\begin{figure}[h]
\includegraphics[width=\columnwidth]{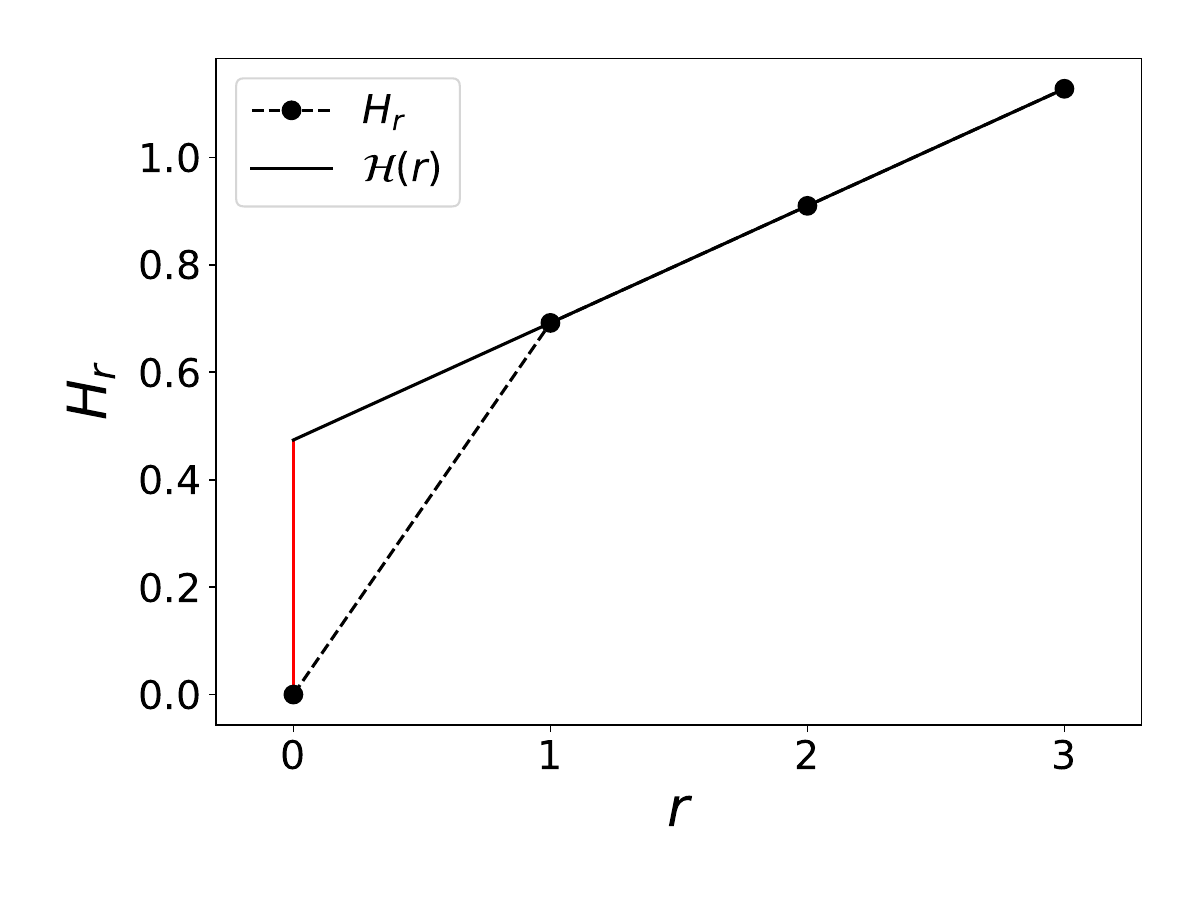}
 \caption{Example of a system with memory $m=1$. Black dots represent the values of $H_r$, and the black solid line shows the linear function $\mathcal{H}(r)$. The red vertical line at $r=0$ indicates the distance between these two curves, with its length corresponding to the value of $\mathcal{G}_0$.}
 \label{fig:example_prop}
\end{figure}

Figure~\ref{fig:example_prop} illustrates an example of a system with memory $m=1$. The black dots represent the computed values of $H_r$, while the black solid line denotes the linear function $\mathcal{H}(r)$. The red vertical segment at $r=0$, indicating the difference between these two curves, corresponds to the value of $\mathcal{G}_0$.

Since the curve $\mathcal{H}(r)$ lies above the graph of $H_r$ at $r=m-1$, this indicates that the actual correlations in the process reduce the overall uncertainty compared to what would be expected if the process exhibited only $(m-1)$th-order dependencies. This is illustrated in the example shown in Fig.~\ref{fig:example}, which depicts a process with memory $m=3$. In panel (a), the value of $\mathcal{G}_2$ corresponds to the vertical distance between the solid line representing $\mathcal{H}(r)$ and the dashed line representing $H_r$ at $r=2$.

All the properties discussed above highlight that the predictability gain is a powerful and informative tool for analyzing dependencies within a stochastic process. It allows for a step-by-step quantification of the strength of temporal correlations, offering a clear and intuitive interpretation.

\subsection{Hypothesis testing}
\label{sec:hyp_test}

The definition of the memory of a process given in Eq.~\eqref{eq:m} naturally raises the question of whether this condition can be simplified. For instance, if one could prove that $\mathcal{G}_u=0$ cannot occur for any $u<m$, then estimating the memory would reduce to finding the first value of $u$ for which the predictability gain vanishes. However, as we show in Appendix~\ref{sec:AppB}, it is indeed possible for a process with memory $m\geq 2$ to have $\mathcal{G}_u$ equal to zero for some values of $u<m-1$. Moreover, for general values of $m$ and $L$, predicting when such occurrences arise is not straightforward. Consequently, determining the memory of a process requires examining all values of $\mathcal{G}_u$ for $u\geq 0$.

To address this difficulty, we propose an algorithm to determine the unknown memory of a process through a sequence of hypothesis tests. 

We start by defining the global null hypothesis $\mathbf{N}^{(\eta)}$ that the process has memory $\eta \geq 0$ 
\begin{equation}
\mathbf{N}^{(\eta)} = \bigcap_{u=\eta}^{\infty} \mathbf{N}_{u}, 
\label{eq:general_null}
\end{equation}
where $\mathbf{N}_{u}$ is the hypothesis that the $(u+1)$th-order transition probabilities can be reduced to $u$th-order ones without loss of information:
\begin{equation}
\mathbf{N}_{u}: p(\beta_{i_{u+1}}|\beta_{i_0},\ldots,\beta_{i_{u}})=p(\beta_{i_{u+1}}|\beta_{i_1},\ldots,\beta_{i_{u}}), 
\label{eq:null}
\end{equation}
for all $i_0,\ldots,i_{u+1}$.

The algorithm begins by stating the null hypothesis $\mathbf{N}^{(0)}$ that the process has memory $0$, against the alternative that the process has memory larger than $0$. We use the predictability gain $\mathcal{G}_u$ as a test statistic. If $\mathcal{G}_u > 0$ for some $u$, we reject $\mathbf{N}_{u}$, which implies that the global hypothesis $\mathbf{N}^{(0)}$ is also rejected. We then proceed to test the next global hypothesis $\mathbf{N}^{(1)}$. The procedure continues by increasing $\eta$ until we fail to reject $\mathbf{N}^{(\eta)}$. The smallest such value is then taken as the memory estimate $m=\eta$. If none are accepted, we conclude that the process cannot be characterized as a finite-memory Markov chain~\cite{PhysRevE.76.011106}. This algorithm is consistent with the definition in Eq.~\eqref{eq:m}.

It is important to note that this procedure assumes exact knowledge of transition probabilities or access to infinite data. In practical situations, this is not the case. Nonetheless, the algorithm provides a theoretical foundation for the memory estimation method we present in the Section~\ref{sec:memory_est}, which is designed for application to finite data sequences.

\section{Memory estimation}
\label{sec:memory_est}

To apply the general memory estimation method outlined in Section~\ref{sec:hyp_test} to real-world data, it must be adapted to practical scenarios where the only available information is a finite data sequence $S$ of length $N$, obtained from consecutive realizations of the random variable $X$.
Since our approach relies on the predictability gain as a test statistic, the first step is to explore how this quantity can be estimated from data. While the estimation of entropy is well-studied~\cite{paninski}, estimating conditional relative entropy or Kullback–Leibler divergence is more challenging, although recent advances have been made~\cite{PhysRevE.109.024305,piga2023bayesian}. For this reason, we will base our estimation of the predictability gain on Eq.~\eqref{eq:pred_gain}, which expresses it in terms of block entropies.

In general, a numerical procedure that approximates the true value of a quantity $a$
based on data is known as an estimator, and is denoted by $\hat{a}$.

Thus, we can estimate the predictability gain given the sequence $S$ as
\begin{equation}
\hat{\mathcal{G}}_u[S] = -\left( \hat{H}_{u+2}[S]-2\hat{H}_{u+1}[S]+\hat{H}_{u}[S] \right),
\label{eq:pred_gain_est}
\end{equation}
where $\hat{H}_r[S]$ represents the estimator of the block entropy of size $r$ applied to $S$, and we recall the convention $\hat{H}_{0}[S]=0 $.

It is well-known that estimating entropy is a challenging task, and numerous estimators exist in the literature~\cite{MM, CS, grassberger, bonachela, shr, chao, grassberger2, contreras}. Here, we will use the NSB entropy estimator~\cite{nsb,nsb2,nsb3}, which has been shown to perform well for correlated sequences~\cite{juan2}. 

To estimate $H_r$ using this method, we first need to group $S$ in overlapping blocks of size $r$ and count the number of times $n(i_1,\ldots,i_r)$ the block $(\beta_{i_1},\ldots,\beta_{i_r})$ occurs in $S$, for $i_1,\ldots,i_r=0,\ldots,L-1$. 
Notice that the number of possible blocks grows exponentially with the block size as $L^r$, but the number of overlapping blocks in the sequences decreases as $N-r+1$.
Given that the performance of all entropy estimators diminishes in the undersampled regime, this implies that there exists a certain $r_{\text{max}}$ up to which we can reliably estimate the block entropy. We will take this value to be
\begin{equation}
r_{\text{max}} = \left \lfloor{\dfrac{\ln(N)}{\ln(L)}}\right \rfloor.
\label{eq:rmax}
\end{equation}
This means that Eq.~\eqref{eq:pred_gain_est} is valid up to $u_{\text{max}}=r_{\text{max}}-2$.

For our goal of memory estimation, we will also need to estimate the probability of occurrence of each block of length $r$ as
\begin{equation}
 \hat{p}(\beta_{i_1},\ldots,\beta_{i_r}) = \dfrac{n(i_1,\ldots,i_r)}{N-r+1},
 \label{eq:prob_est_block}
\end{equation}
as well as the $u$th-order transition probabilities as
\begin{equation}
 \hat{p}(\beta_{i_{u+1}}|\beta_{i_1},\ldots,\beta_{i_u}) = \dfrac{n(i_1,\ldots,i_{u+1})}{\sum_{l=0}^{L-1}n(i_1,\ldots,i_{u},l)}.
 \label{eq:trans_prob_est}
\end{equation}

\subsection{Method}
\label{sec:method}

As described in Section~\ref{sec:hyp_test}, the method to determine whether a process has memory $\eta$ involves performing a sequence of hypothesis tests for increasing values of $u$, starting at $u = \eta$.
In practice, since we can only reliably estimate the predictability gain up to $u_{\text{max}}$, the tests are limited to this range.

Additionally, due to the statistical fluctuations inherent in estimating the predictability gain via Eq.~\eqref{eq:pred_gain_est}, it is not sufficient to reject the global hypothesis $\mathbf{N}^{(\eta)}$ solely based on the condition $\hat{\mathcal{G}}_u>0$ for some $\eta\leq u \leq u_{\text{max}}$. A more robust strategy is to compare the observed values of $\hat{\mathcal{G}}_u$ with those expected under the assumption that $\mathbf{N}^{(\eta)}$ holds.
This comparison can be carried out by computing the p-value $q_u^{(\eta)}$, defined as~\cite{pvalue}
\begin{equation}
q_u^{(\eta)} = P\left(\hat{\mathcal{G}}_u[\tilde{S}]\geq\hat{\mathcal{G}}_u[S]\,|\,\tilde{S}\in \mathbf{N}^{(\eta)}\right),
\label{eq:q_u}
\end{equation}
where $\tilde{S}\in \mathbf{N}^{(\eta)}$ indicates that $\tilde{S}$ is a sequence generated by a process with memory $\eta$.
The p-value given by Eq.~\eqref{eq:q_u} quantifies the probability of observing a value of the predictability gain at least as extreme as the one obtained from the original sequence $S$, assuming that the null hypothesis $\mathbf{N}^{(\eta)}$ holds.

Computing $q_u^{(\eta)}$ exactly requires knowledge of the distribution of $\hat{\mathcal{G}}_u$ when acting on sequences of memory $\eta$, which is not straightforward. However, non-parametric methods exist for approximating such p-values~\cite{https://doi.org/10.1002/widm.1054}.
In particular, we adopt the bootstrap method, a resampling technique that generates synthetic samples from the observed data~\cite{mackinnon2009bootstrap}.
In the context of hypothesis testing, this technique provides an empirical approximation of the distribution of the test statistic under the null hypothesis. This allows us to estimate p-values by comparing the observed value of the statistic to the distribution obtained from the resampled data.

To apply the bootstrap method in the context of memory estimation, we assume the null hypothesis $\mathbf{N}^{(\eta)}$ which states that the sequence $S$ was generated by a process with memory $\eta\geq 0$. Under this assumption, we estimate the probabilities of the blocks of size $\eta$ and the $\eta$th-order transition probabilities using Eqs.~\eqref{eq:prob_est_block} and~\eqref{eq:trans_prob_est}, respectively.
These estimated probabilities are then used to generate $K$ synthetic sequences $\tilde{S}_1^{\eta},\ldots,\tilde{S}_K^{\eta}$, each of length $N$, which constitute the bootstrap samples.
By construction, these sequences have memory $\eta$.

For each bootstrap sample, we compute $\hat{\mathcal{G}}_u[\tilde{S}_k^{\eta}]$, with $k=1,\ldots,K$. Then, the p-value defined in Eq.~\eqref{eq:q_u} can be estimated empirically as
\begin{equation}
\hat{q}_u^{(\eta)} = \dfrac{1}{K} \sum_{k=1}^K I\left(\hat{\mathcal{G}}_u[\tilde{S}_k^{\eta}]\geq\hat{\mathcal{G}}_u[S]\right),
\label{eq:estimated_q_u}
\end{equation}
where $I(A)$ is the indicator function, that yields $1$ if $A$ is true and $0$ otherwise.
Repeating this procedure for $\eta\leq u \leq u_{\text{max}}$, we obtain the full set of p-values associated with the global null hypothesis $\mathbf{N}^{(\eta)}$.

When conducting multiple hypothesis tests, directly comparing each p-value to a fixed threshold $\alpha$ increases the overall probability of committing a type I error
(incorrectly rejecting a true null hypothesis). 
In fact, if the tests are independent, the probability of making at least one type I error across $M$ tests is given by $1-(1-\alpha)^M$~\cite{sheskin2003handbook}.
For instance, if $\alpha=0.05$ and $M=5$, this probability exceeds $22\%$.
In our setting, the global null hypothesis $\mathbf{N}^{(\eta)}$ is composed of $M^{(\eta)}=u_{\text{max}}-\eta+1$ null hypotheses.

Inflation of type-I error in multiple testing scenarios is often addressed by adjusting individual p-values~\cite{noble2009does}. In this work, however, we take an alternative approach: instead of correcting each p-value, we combine them into a single statistic that can be directly compared to the significance threshold $\alpha$. This approach ensures that the overall type I error rate for the global null hypothesis remains controlled at the desired level.

A variety of methods for combining p-values have been proposed in the literature~\cite{LOUGHIN2004467,doi:10.1073/pnas.1814092116}. Here, we adopt Fisher’s method~\cite{fisher1970statistical} to test the global null hypothesis $\mathbf{N}^{(\eta)}$. The combined p-value is computed as (see Appendix~\ref{app:fisher} for details).
\begin{equation}
\hat{q}^{(\eta)} = z^{(\eta)}\sum_{j=0}^{u_{\text{max}}-\eta}\dfrac{\left(-\ln(z^{(\eta)})\right)^j}{j!},
\end{equation}
with
\begin{equation}
z^{(\eta)} = \prod_{u=\eta}^{u_{\text{max}}} \hat{q}_u^{(\eta)}.
\end{equation}

We then define the Predictability Gain (PG)  memory estimator as
\begin{equation}
\hat{m}^{\text{\tiny{PG}}} = \text{min}\left(\left\lbrace \eta:\, \hat{q}^{(\eta)}>\alpha,\ 0\leq \eta \leq u_{\text{max}} \right\rbrace \right),
\label{eq:m_est}
\end{equation}
which is the smallest value of $\eta$ for which we fail to reject the null hypothesis $\mathbf{N}^{(\eta)}$.
If all hypotheses up to $u_{\text{max}}$ are rejected, we conclude that the process cannot be represented as a Markov chain of order less than or equal to $u_{\text{max}}$.
However, due to the properties and interpretation of the predictability gain discussed in Section~\ref{sec:properties}, valuable insights into the system’s structure and temporal dependencies can still be obtained.

It is worth noting that the error probability of this estimator depends on the true memory of the process. If the sequence is iid (i.e., $m=0$), the only error occurs if $\mathbf{N}^{(0)}$ is incorrectly rejected, which happens with probability $\alpha$. For processes with memory $m\geq 1$, there are two sources of error: falsely accepting $\mathbf{N}^{(\eta)}$ for some $\eta < m$, or incorrectly rejecting $\mathbf{N}^{(m)}$, which again occurs with probability $\alpha$. Therefore, the overall error rate can exceed $\alpha$ in the general case.

Before applying this estimator to real data, we will validate its performance using synthetic sequences with known memory to assess its accuracy and robustness.

\subsection{Simulations}

We compare our proposed memory-estimation method with two widely used alternatives: AIC and BIC, which aim to balance model fit and complexity. Both criteria start from the log-likelihood, rewarding models that closely reproduce the observed data, and then add a penalty—different for each—that discourages overfitting by favoring more parsimonious models with fewer parameters. In both cases, lower values indicate the preferred model.

Specifically, for a Markov model of order $\eta \geq 0$ with $L$ possible outcomes, the log-likelihood function $\hat{l}(\eta)$ associated to an observed sequence $S$ is given by
\begin{equation}
\hat{l}(\eta) = \sum_{i_1,\ldots,i_{\eta+1}=0}^{L-1} n(i_1\ldots i_{\eta+1})\ln(\hat{p}(\beta_{i_{\eta+1}}|\beta_{i_{1}},\ldots,\beta_{i_{\eta}})).
\end{equation}
Then, the AIC function reads
\begin{equation}
A(\eta) = -2\hat{l}(\eta)+2L^{\eta}(L-1),
\label{eq:aic}
\end{equation}
whereas the BIC function is defined as
\begin{equation}
B(\eta) = -2\hat{l}(\eta)+L^{\eta}(L-1)\ln(N).
\label{eq:bic}
\end{equation}

Thus, the proposed AIC and BIC memory estimators are the values of $\eta$ that minimize Eqs.~\eqref{eq:aic} and~\eqref{eq:bic}, respectively:
\begin{equation}
\begin{split}
\hat{m}^{\text{\tiny{AIC}}} &= \mathop{\arg \min}\limits_{0\leq \eta \leq u_{\text{max}}} A(\eta), \\
\hat{m}^{\text{\tiny{BIC}}} &= \mathop{\arg \min}\limits_{0\leq \eta \leq u_{\text{max}}} B(\eta).
\end{split}
\end{equation}

To assess the performance of the PG, AIC, and BIC memory estimators, we consider binary stochastic processes ($\beta_0=0$, $\beta_1=1$), with memory values $m=0,1,2,3,4$. For each value of $m$, we generate $J$ distinct processes, each one assigning transition probabilities of order $m$, $p(0|\beta_{i_1},\ldots,\beta_{i_m})$,  for $i_1,\ldots,i_m=0,1$, randomly drawn from a uniform distribution on $[0,1]$. The complementary probabilities are then set as $p(1|\beta_{i_1},\ldots,\beta_{i_m})=1-p(0|\beta_{i_1},\ldots,\beta_{i_m})$.

For each of these $J$ processes, we simulate a sequence of length $N$, and estimate the memory order using the three methods. Estimator accuracy is evaluated as the proportion of sequences for which the inferred memory matches the true generating order.

It can be shown~\cite{DeGregorio2025} that, when the transition probabilities of a binary Markov process are randomly drawn from a uniform distribution, the median of $\mathcal{G}_0$ is approximately $0.04$. This implies that half of the generated processes exhibit a predictability gain below $6\%$ of the theoretical maximum ($\sim 0.69$).
Motivated by this observation, and by our focus on correctly identifying memory in cases where underestimation would lead to a substantial loss of information, we restrict our analysis to processes satisfying $\mathcal{G}_{m-1}>0.04$, if $m\geq 1$.

For the computation of the proposed PG memory estimator in Eq.~\eqref{eq:m_est}, we fix the significance level at $\alpha=0.05$ and use $K=2000$ bootstrap samples to estimate the p-values in Eq.\eqref{eq:estimated_q_u}.

Table~\ref{tab:m_estimators} reports the accuracy of the three memory estimators for sequence lengths $N=100$, $200$, and $300$, based on $J=500$ independently sampled sets of transition probabilities.
The code used to estimate predictability gain and memory values with the PG method and to generate the simulated data for Table~\ref{tab:m_estimators} is publicly available at Ref.~\cite{code}.

\begin{table}[htb]
\centering
\small
\squeezetable
\begin{tabular*}{\columnwidth}{@{\extracolsep{\fill}} | l | *{5}{c} | *{5}{c} | *{5}{c} |}
\hline
 \multicolumn{1}{|c|}{$N$} & \multicolumn{5}{c|}{$100$} & \multicolumn{5}{c|}{$200$} & \multicolumn{5}{c|}{$300$} \\
\cline{1-6}\cline{7-11}\cline{12-16}
\multicolumn{1}{|c|}{$m$} & $0$ & $1$ & $2$ & $3$ & $4$ & $0$ & $1$ & $2$ & $3$ & $4$
& $0$ & $1$ & $2$ & $3$ & $4$ \\
\hline
\hline
PG & $\mathbf{95}$ & $\mathbf{87}$ & $\mathbf{74}$ & $62$ & $60$ &
  $\mathbf{96}$ & $\mathbf{94}$ & $\mathbf{89}$ & $\mathbf{88}$ & $\mathbf{79}$ &
  $\mathbf{95}$ & $\mathbf{93}$ & $\mathbf{93}$ & $\mathbf{92}$ & $\mathbf{86}$ \\
AIC & $42$ & $46$ & $55$ & $\mathbf{64}$ & $\mathbf{62}$ &
  $34$ & $38$ & $45$ & $56$ & $62$ &
  $29$ & $27$ & $33$ & $39$ & $47$ \\
BIC & $83$ & $\mathbf{87}$ & $\mathbf{74}$ & $45$ & $4$ &
  $85$ & $86$ & $86$ & $76$ & $31$ &
  $80$ & $83$ & $88$ & $87$ & $57$ \\
\hline
\end{tabular*}
\caption{Percentage of correctly estimated memory values for the PG (Predictability Gain), AIC, and BIC methods across $J=500$ binary sequences ($L=2$) of lengths $N=100$, $200$, and $300$. The sequences were generated using randomly sampled transition probabilities for memory orders $m=0,1,2,3$, and $4$. The highest accuracy in each case is indicated in bold.}
\label{tab:m_estimators}
\end{table}

As shown in Table~\ref{tab:m_estimators}, the PG estimator achieves higher accuracy than both AIC and BIC for nearly all combinations of $N$ and $m$. The only exception occurs for $N=100$ and $m=3,4$, where AIC performs slightly better. However, the accuracy of AIC decreases noticeably with increasing $N$ across all memory orders. Although this behavior may seem counterintuitive, it is consistent with previous findings that AIC tends to overestimate model complexity~\cite{dorea2014simulation}. As $N$ increases, so does the maximum allowed memory $u_{\text{max}}$, which we set to $4, 5$, and $6$ for $N=100$, $200$, and $300$, respectively. Since AIC has a preference for higher memory values, this results in lower accuracy as $u_{\text{max}}$ grows. Interestingly, when we fix $u_{\text{max}}=4$ for all values of $N$, AIC's performance improves significantly, reaching $88\%$ accuracy for $N=300$ and $m=4$. This behavior reflects the well-known inconsistency of AIC~\cite{https://doi.org/10.2307/2347338}.

The BIC estimator, on the other hand, is known to be consistent~\cite{10.1214/aos/1015957472}, which is in line with our results: its accuracy remains relatively stable as $N$ increases, unlike AIC. However, BIC tends to favor smaller memory values and is known to perform poorly when the sample size is limited~\cite{wea99}. This is evident in its low accuracy for $N=100$ and higher memory values. Although its performance improves with longer sequences, it still falls short of the accuracy achieved by the PG estimator.

Overall, Table~\ref{tab:m_estimators} demonstrates that the PG estimator is notably more reliable than both AIC and BIC when applied to binary sequences with memory $m=0,1,2,3,4$. It not only achieves higher accuracy but also exhibits consistent performance improvements as the sequence length increases.

Unlike AIC and BIC, which are designed to always return a model within the candidate range regardless of whether the true memory falls within it, the PG estimator offers a key advantage: it can flag situations where the assumed range may be insufficient. Specifically, if none of the computed p-values exceed the significance threshold, the PG estimator indicates that no suitable memory value was detected within the tested range.
However, the performance of the PG estimator can be influenced by the choice of significance level $\alpha$ and the number of bootstrap samples used for p-value estimation. Additionally, the computational cost of generating these samples can become substantial, particularly for long sequences.

Although numerous memory estimators have been proposed in the literature~\cite{PhysRevE.76.011106,1056936,https://doi.org/10.1002/cjs.11225}, special attention should be given to the method introduced in Ref.~\cite{PAPAPETROU20131593}. The function analyzed in that work—referred to by the authors as conditional mutual information—is in fact equivalent to the predictability gain used in our study. Nonetheless, there are two key differences between their approach and ours.

First, their estimator defines the memory as the smallest value of $u$ for which the predictability gain vanishes. However, this definition may lead to underestimation of the true memory, since the predictability gain can drop to zero even before the actual memory order is reached, as shown in Appendix~\ref{sec:AppB}. In contrast, our method avoids this limitation by evaluating the full sequence of predictability gain values and testing their statistical significance.

Second, their hypothesis testing procedure relies on permutation tests~\cite{doi:10.1080/01621459.1990.10474929}, where the observed sequence is shuffled to generate surrogate data. By comparing the empirical values of $\hat{\mathcal{G}}_u$ with those obtained from these surrogates, the memory is defined as the minimum $u$ for which the corresponding p-value exceeds a predefined threshold $\alpha$. However, since the surrogate sequences are fully randomized and therefore uncorrelated, the null hypotheses being tested effectively assume memoryless (iid) behavior. As a result, it is unclear whether failing to reject the null at a given $u$ genuinely supports the conclusion that the process has memory $u$, rather than simply indicating a lack of sufficient evidence to distinguish it from a memoryless one.

Given these differences, our approach is more robust and offers a clearer interpretation of the results.

In a related context, it is worth mentioning that alternative frameworks for selecting the effective past relevant for the evolution of a system rely on uniform and non-uniform embedding schemes~\cite{PhysRevE.91.032904}.

\section{Precipitation sequences}
\label{sec:prec_seq}

It is generally recognized that the occurrence of daily precipitation can be reasonably approximated by a first-order Markov process, where the probability of rain on a given day depends solely on whether it rained the previous day~\cite{WR013i006p00949}.
However, several studies have proposed extending the model to include higher-order dependencies, effectively increasing the memory length of the Markov process~\cite{schoof2008proper}. This allows the model to account for cumulative effects from multiple preceding days, thereby better capturing the influence of long-lived weather systems.
Moreover, the memory structure of precipitation sequences has been shown to vary with both geographic location and season~\cite{https://doi.org/10.1002/joc.7175}. In some regions, memory length increases during particular times of the year due to recurring atmospheric patterns, whereas in others it remains relatively stable year-round.

Beyond estimating the memory order of precipitation sequences for specific locations and months, a complementary objective is to quantify the strength of temporal correlations between precipitation events across multiple days. This provides additional insight into the underlying dynamics and can help identify regimes where longer-term dependencies play a more significant role.

In this section, we investigate memory effects in sequences of daily precipitation across the contiguous United States. Our goal is to estimate the memory order of these sequences, quantify the strength of the first-order correlations and analyze how they fluctuate across different seasons and regions.

\subsection{Data}

Daily precipitation data were obtained from the Global Historical Climatology Network Daily dataset~\cite{menne2012overview}, which compiles weather observations from a large network of meteorological stations worldwide. The dataset includes daily totals of recorded precipitation at each station. For the purposes of this study, we apply a binary classification to each day based on the presence or absence of precipitation. Following common practice~\cite{https://doi.org/10.1002/joc.7175}, a day is classified as dry and assigned $\beta_0=0$ if the total precipitation is less than $0.1$ mm; otherwise, it is considered wet and assigned $\beta_1=1$.
This binary ($L=2$) discretization allows us to focus on the occurrence of precipitation events rather than their magnitude, thereby simplifying the analysis of temporal patterns in rainfall occurrence.

We focus on stations located in the contiguous United States that report daily precipitation data spanning the period from January 1, 1990, to December 31, 2020. To capture the spatio-temporal variation of memory effects, we organize the data by both station and calendar month.

For each station, we construct a collection of sequences by separating the data month by month. For instance, at a given station where the data in the period considered is complete, we define a set $\lbrace S_1,\ldots,S_{31} \rbrace$ for January, where each sequence $S_v$ corresponds to the binary precipitation data for January of year $1990+v-1$. Specifically, $S_1$ contains data for January 1990, $S_2$ for January 1991, and so on up to $S_{31}$ for January 2020. The same procedure is applied to each subsequent month, yielding up to $12$ monthly sets of sequences per station.

More generally, for each station and month, we define a set of binary sequences $\lbrace S \rbrace = \lbrace S_1,\ldots,S_{V} \rbrace$, where $S_v$ is of length $N_v$. If data availability is complete, then $V=31$. However, due to gaps or limited observation periods, some station-month combinations may have different number of sequences.

To ensure sufficient data quality for statistical analysis, we discard any station-month pair $\lbrace S \rbrace$ for which the total number of available days is less than $300$. 
In our analysis we include a total of approximately $8000$ stations, although the exact number may vary slightly from month to month depending on data availability.

\subsection{Memory}

Using the set of sequences $\lbrace S \rbrace$, we estimate the predictability gain and the memory of the process following the procedure described in Section~\ref{sec:method}, with the number of bootstrap samples set to $K=2000$ and a significance level of $\alpha=0.05$.

\begin{figure}[h]
 \centering

 \begin{minipage}{\linewidth}
 \begin{minipage}{0.05\linewidth}
  \raggedright \textbf{(a)}
 \end{minipage}
 \begin{minipage}{0.9\linewidth}
  \includegraphics[width=\linewidth]{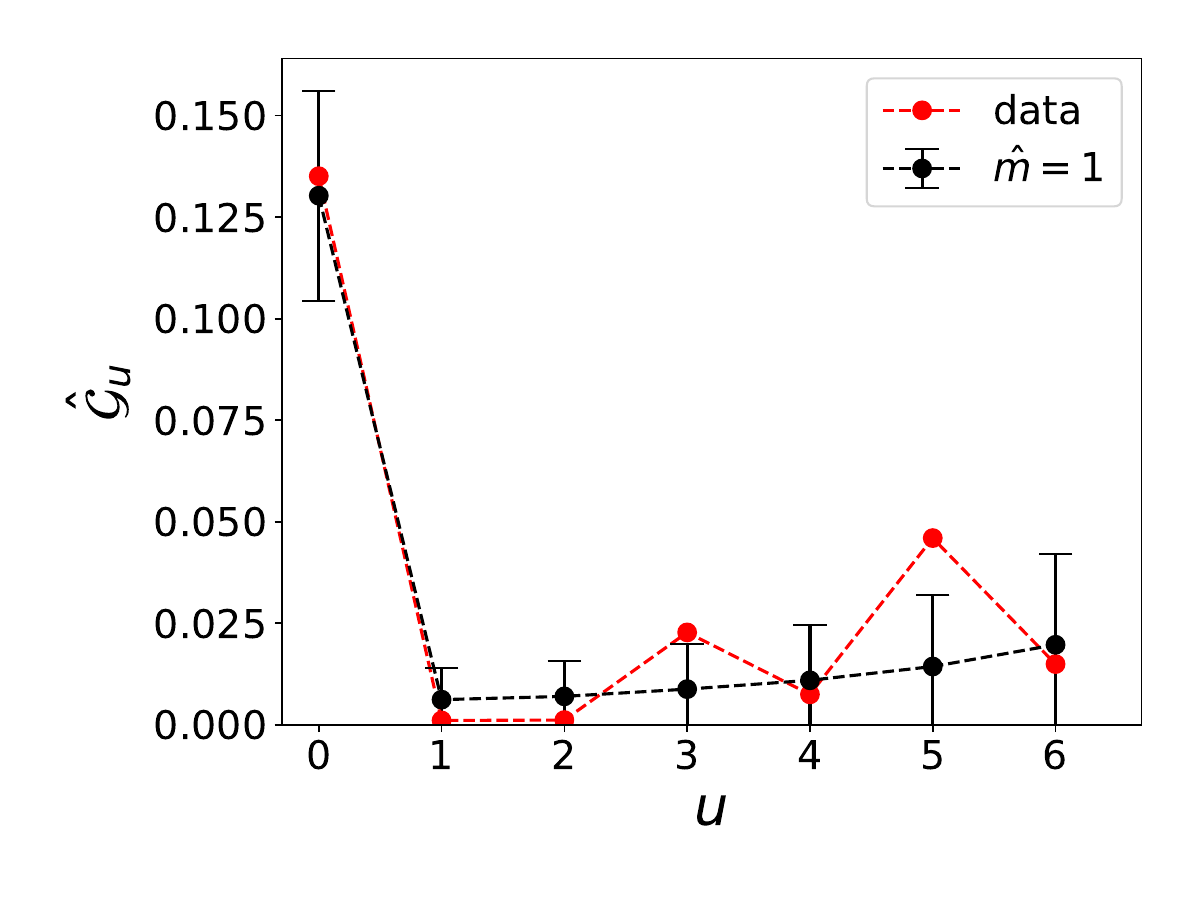}
 \end{minipage}
 \end{minipage}

 \vspace{2ex}

 \begin{minipage}{\linewidth}
 \begin{minipage}{0.05\linewidth}
  \raggedright \textbf{(b)}
 \end{minipage}
 \begin{minipage}{0.9\linewidth}
  \includegraphics[width=\linewidth]{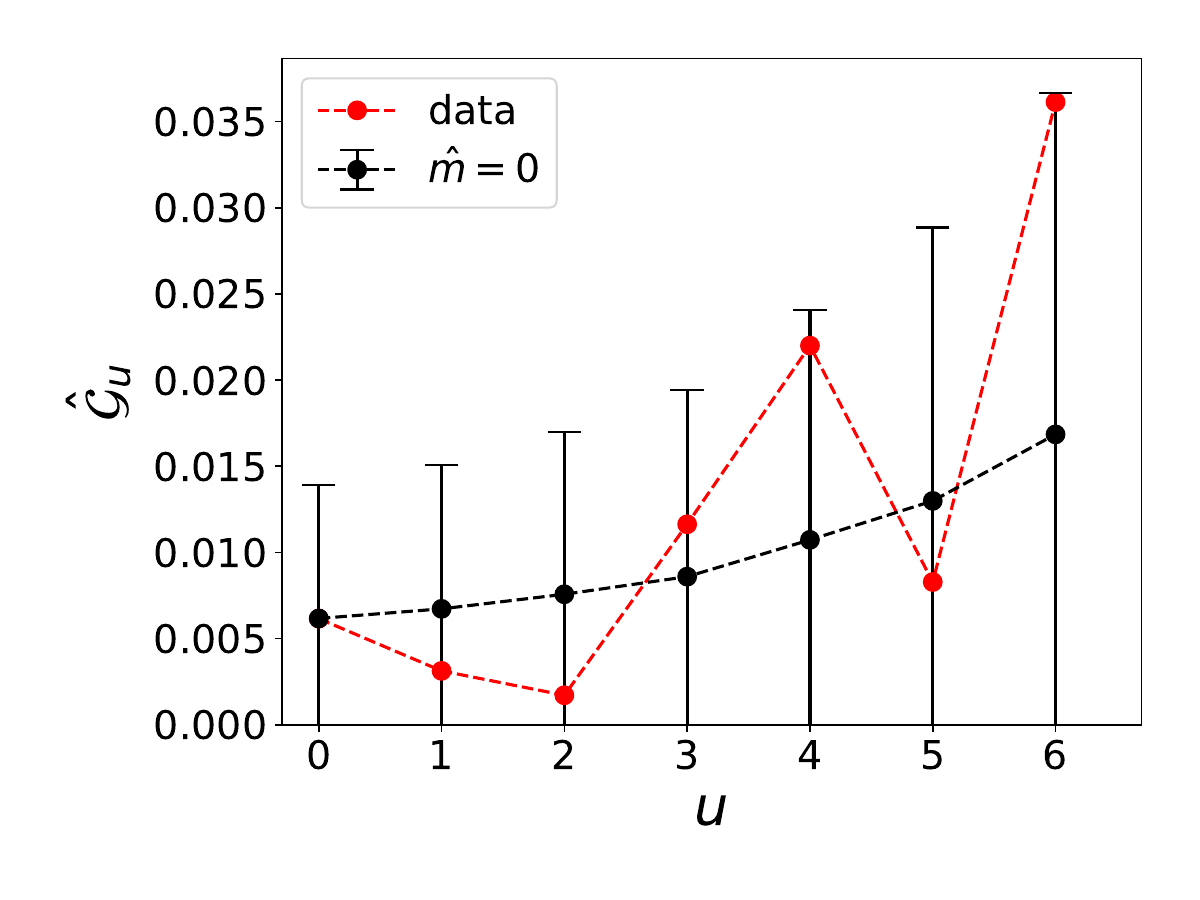}
 \end{minipage}
 \end{minipage}

 \caption{Estimated predictability gain for a station located in Coos Bay, Oregon, for January (a) and August (b), shown in red. The mean and sample standard deviation, shown in black, are computed from $K=2000$ bootstrap samples generated numerically based on the estimated memory values: $\hat{m}^{\text{\tiny{PG}}}=1$ for (a) and $\hat{m}^{\text{\tiny{PG}}}=0$ for (b).}
 \label{fig:rain_example}
\end{figure}

Fig.~\ref{fig:rain_example} presents two examples of the estimated predictability gain (shown in red) for a station located in Coos Bay, Oregon, corresponding to January (panel a) and August (panel b). The estimated memory values for these cases are $1$ and $0$, respectively.
For comparison, the mean predictability gain $\bar{\mathcal{G}}_u$ and the sample standard deviation $s_u$ are shown in black. These quantities are computed from $K$ bootstrap samples $\lbrace \tilde{S} \rbrace_1,\ldots,\lbrace \tilde{S} \rbrace_K$, where each sample $\lbrace \tilde{S} \rbrace_k$ is of the same size as the original set $\lbrace S \rbrace$ and is generated based on the estimated memory for the corresponding case.
The mean and standard deviation are defined as
\begin{equation}
\bar{\mathcal{G}}_u = \dfrac{1}{K} \sum_{k=1}^K \hat{\mathcal{G}}_u[\lbrace \tilde{S} \rbrace_k],
\end{equation}
and
\begin{equation}
s_u = \sqrt{\dfrac{1}{K-1}\sum_{k=1}^K \left(\hat{\mathcal{G}}_u[\lbrace \tilde{S} \rbrace_k]-\bar{\mathcal{G}}_u \right)^2}.
\end{equation}

In both panels of Fig.~\ref{fig:rain_example}, the red curve (data-based estimate of predictability gain) is in good agreement with the black curve (model-based expectation), indicating that the estimated predictability gain falls well within the expected range under the fitted memory model.

The results shown in Fig.~\ref{fig:rain_example} illustrate that the memory of precipitation sequences varies with the time of year. This seasonal dependence is further supported by Table~\ref{table:precip}, which reports the monthly percentages of stations with estimated memory values $m=0,1,2,3,4$.

\begin{table}[h]
\centering
\begin{tabular}{|c|c c c c c|}
\hline
\diagbox{\textbf{Month}}{\textbf{Memory}} & 0 & 1 & 2 & 3 & 4 \\ \hline
1 & 40 & \textbf{53} & 3 & 1 & 3 \\ 
2 & 45 & \textbf{48} & 3 & 1 & 3 \\ 
3 & 30 & \textbf{61} & 5 & 1 & 3 \\ 
4 & 31 & \textbf{62} & 2 & 1 & 3 \\ 
5 & 13 & \textbf{79} & 3 & 1 & 3 \\ 
6 & 33 & \textbf{61} & 2 & 1 & 3 \\ 
7 & 42 & \textbf{52} & 2 & 1 & 3 \\ 
8 & 45 & \textbf{50} & 2 & 1 & 2 \\ 
9 & 23 & \textbf{70} & 3 & 1 & 3 \\ 
10 & 15 & \textbf{75} & 3 & 2 & 5 \\ 
11 & 33 & \textbf{63} & 2 & 1 & 2 \\ 
12 & 34 & \textbf{59} & 4 & 1 & 2 \\ \hline
\end{tabular}
\caption{Percentage of stations with estimated memory values of $0$, $1$, $2$, $3$, and $4$ for each month. The highest frequency is highlighted in bold.}
\label{table:precip}
\end{table}

We observe that, throughout the year, the majority of stations are characterized by memory $1$, with this dominance being particularly pronounced in May, September, and October. Nonetheless, a considerable fraction of stations also exhibit memory $0$, especially in winter (January, February) and summer (July, August), where the distribution between memory $0$ and memory $1$ is more balanced.

When considering each station individually, the majority ($\sim 67\%$) most frequently display a memory value of 1 throughout the
year, in agreement with findings in Ref.~\cite{https://doi.org/10.1002/joc.7175}.

In Appendix~\ref{app:C}, Fig.~\ref{fig:spatial_month} shows the spatial distribution of the estimated memory of the Markov process for each month, starting with December in panel (a) and ending with November in panel (l). The distribution of stations with $\hat{m}=0$ exhibits a clear seasonal variability. For most of the year, these stations are concentrated in the central and eastern United States, whereas during summer months a higher occurrence of uncorrelated patterns emerges in the west, particularly in California.

In contrast, stations with an estimated order of $2$ are primarily located in the Northeast, Southeast, and Ohio Valley~\cite{karl1984regional}. This pattern is particularly evident in panels (a) and (d), which, according to Table~\ref{table:precip}, correspond to the months with the highest occurrence of stations with $\hat{m}=2$.

It should be noted that the $99$th percentile of the calculated values of $\hat{\mathcal{G}}_1$ across all stations and months is $\sim 0.02$, which represents less than $3\%$ of the maximum possible value the predictability can take, according to Proposition~\ref{prop:bounded}. Additionally, Table~\ref{table:precip} shows that memory values larger than $1$ occur only rarely. Therefore, we can conclude that most predictive information is already included in the first-order transition probabilities.

\subsection{First-order correlations}

Markov chains of order $0$ and 
$1$ were found to be highly predominant across all months and stations considered.
Our objective is now to quantify the strength of the first-order
correlations by analyzing the values of $\hat{\mathcal{G}}_0$ and their variability with the first-order transition probabilities $\hat{p}(0|0)$ and $\hat{p}(1|1)$, calculated using Eq.~\eqref{eq:trans_prob_est}.
For station–month pairs with an estimated memory of $0$, we assign $\hat{\mathcal{G}}_0=0$, reflecting the absence of correlations in such cases.

\begin{figure}[h]
\includegraphics[width=\columnwidth]{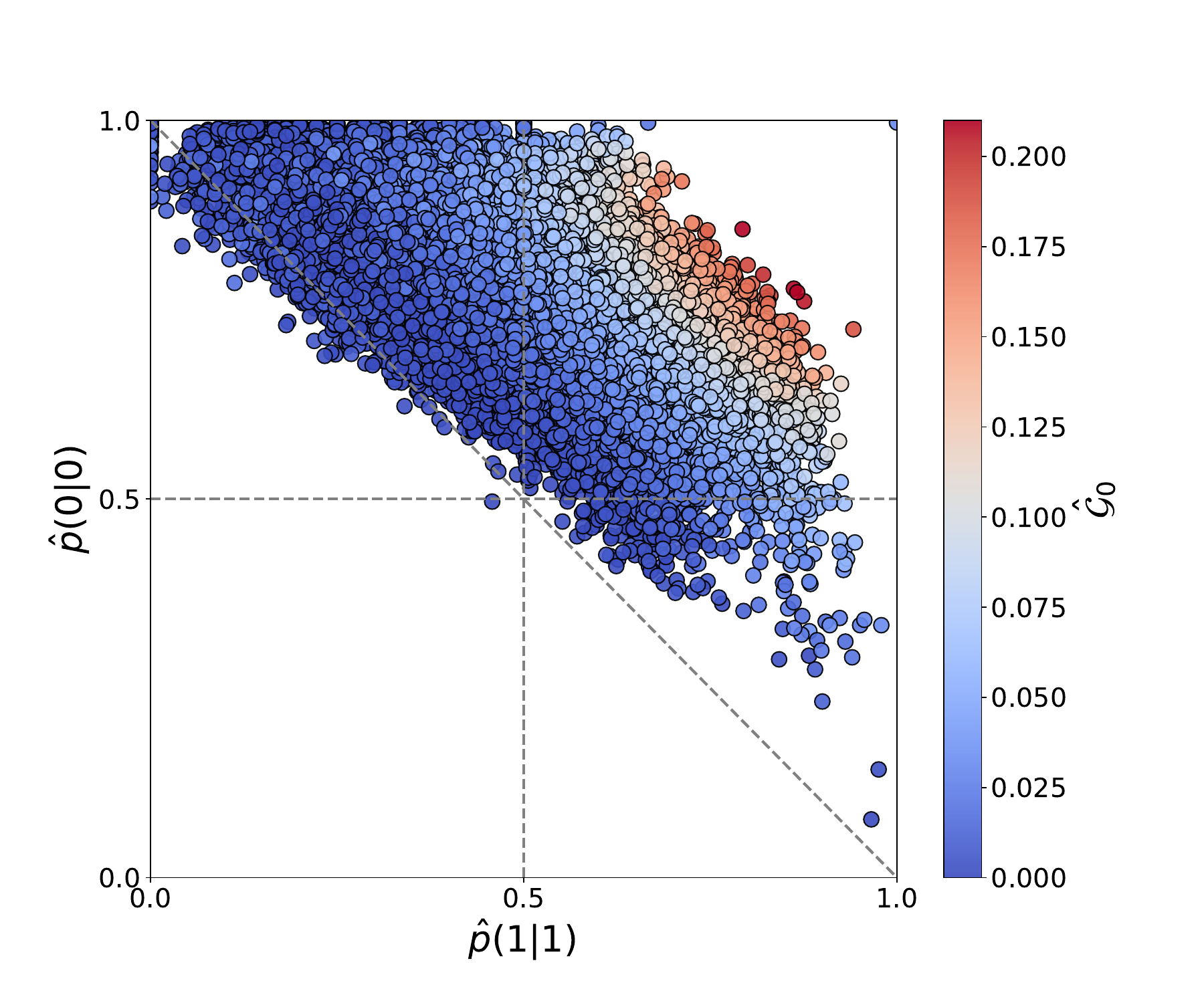}
 \caption{Estimated first-order transition probabilities for each station-month pair are shown as dots, with colors indicating the corresponding value of $\hat{\mathcal{G}}_0$. Dashed lines mark $\hat{p}(0|0)=0.5$, $\hat{p}(1|1)=0.5$, and the diagonal $\hat{p}(0|0)=1-\hat{p}(1|1)$ (iid case).}
 \label{fig:all_G0}
\end{figure}

In Fig.~\ref{fig:all_G0}, the estimated first-order transition probabilities for each station–month pair are shown as dots, with colors indicating the corresponding value of $\hat{\mathcal{G}}_0$.
For reference, dashed lines indicate the special cases $\hat{p}(0|0)=0.5$ and $\hat{p}(1|1)=0.5$, as well as the diagonal $\hat{p}(0|0)=1-\hat{p}(1|1)$, which corresponds to the iid situation.

We observe that the vast majority ($> 99\%$) of station-month pairs present $\hat{p}(0|0)>0.5$, indicating that throughout the territory and across all months, a dry day is more likely to be followed by another dry day. In contrast, only $30\%$ of the cases exhibit $\hat{p}(1|1)>0.5$, reflecting a predominant tendency for a rainy day to be followed by a dry one.

Additionally, it can be seen in Fig.~\ref{fig:all_G0} that for fixed values of $\hat{p}(0|0)$, $\hat{\mathcal{G}}_0$ increases with $\hat{p}(1|1)$. A similar behavior is observed when fixing $\hat{p}(1|1)$ and increasing $\hat{p}(0|0)$. This monotonic relationship is supported by the partial Spearman correlation coefficients~\cite{10.1093/mnras/199.4.1119}, which are $\sim 0.9$ in both cases, with p-values $<0.001$.
These results reveal a strong positive association between $\hat{\mathcal{G}}_0$ and the transition probabilities, showing that higher persistence in either wet or dry conditions leads to stronger first-order correlations.

\subsection{Seasonal and regional variability}

For each station, we calculate the seasonal averages of $\hat{\mathcal{G}}_0$. As a reference, we find that the three largest mean values ($\sim 0.17$, corresponding to approximately $25\%$ of the maximum possible value, according to Proposition~\ref{prop:bounded}) are observed during autumn in the northwestern region of the country, particularly in the states of Oregon and Washington.

\begin{figure}[h]
\includegraphics[width=\columnwidth]{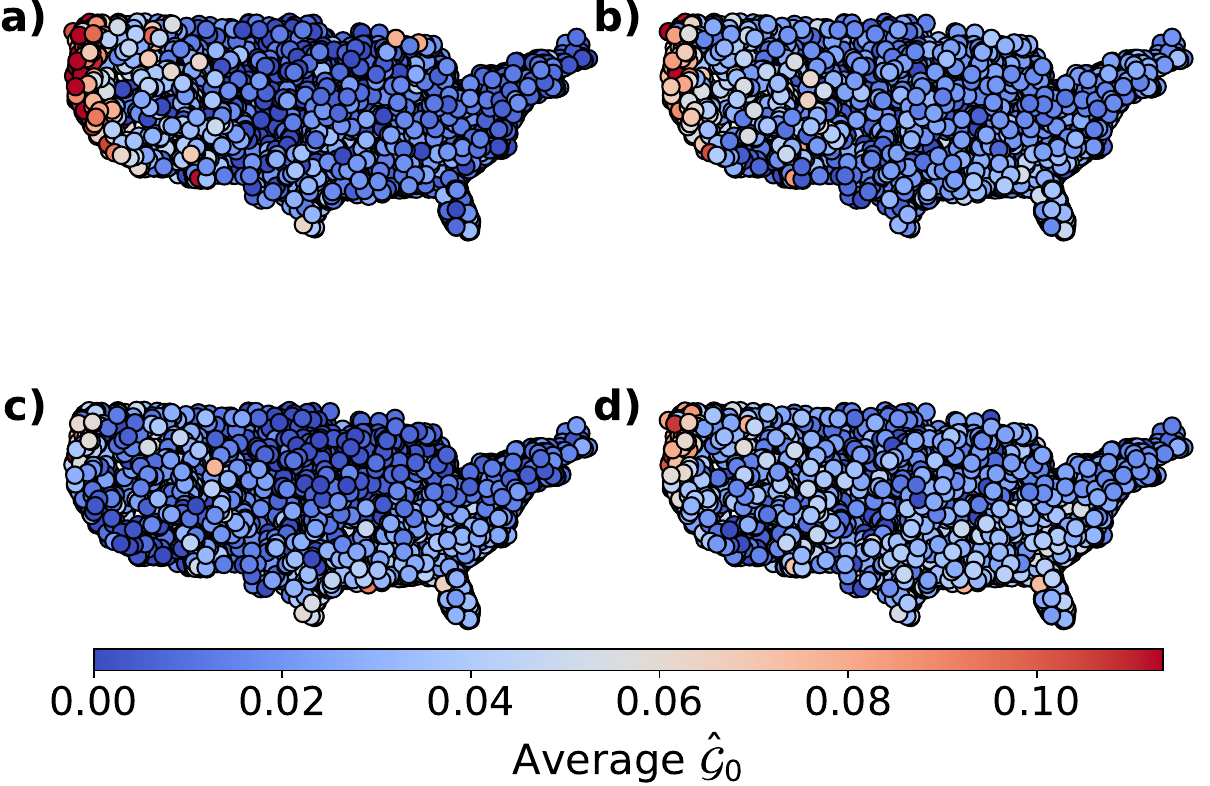}
 \caption{Seasonal averages of $\hat{\mathcal{G}}_0$ for each station. Winter (December–February) in panel (a); spring (March–May) in panel (b); summer (June–August) in panel (c); and autumn (September–November) in panel (d). The colorbar is saturated at the $99$th percentile ($\sim 0.11$) to enhance the visibility of lower values.}
 \label{fig:seasons}
\end{figure}

In Fig.~\ref{fig:seasons}, we present colormaps of the seasonal averages of $\hat{\mathcal{G}}_0$ for each station: winter (December–February) in panel (a), spring (March–May) in panel (b), summer (June–August) in panel (c), and autumn (September–November) in panel (d). 
It should be noted that, in order to enhance the visibility of the lowest values,  the colorbar is saturated at the 99th percentile ($\sim 0.11$).
Correlations above this threshold are still present but are represented uniformly within the uppermost color bin.

Fig.~\ref{fig:seasons} demonstrates how the proposed methodology can uncover differences in temporal correlations across both space and time. For example, correlations are strongest along the West Coast during winter and in the Southeast during summer and spring, while much of the central United States shows comparatively weak values throughout the year.
Transitional seasons (spring and autumn) exhibit intermediate values, with substantial spatial variability. 
Interestingly, when considering which season yields the strongest correlations at each station, autumn dominates with $47\%$ of the cases, while winter, spring, and summer account for only $16\%$, $24\%$, and $13\%$, respectively.

Regionally, panel (a) shows that high winter values of $\hat{\mathcal{G}}_0$ appear in Washington, Oregon, and California. While Washington and Oregon remain among the areas with the strongest correlations during spring and autumn, this pattern weakens in summer. In contrast, correlations in California diminish almost completely during summer, consistent with its pronounced seasonal variability in precipitation~\cite{https://doi.org/10.1029/2008JD010251}.
Comparatively higher values of $\hat{\mathcal{G}}_0$ emerge in the Southeast during summer.
This is supported by one-sided Mann–Whitney tests~\cite{birnbaum1956use}, which indicate that correlations along the West Coast are significantly higher than in the Southeast for all seasons except summer (p-values $< 0.001$).

Given the substantial differences in the strength of correlations observed in the West Coast and Southeast in winter and summer, we now focus on these specific regions and seasons.

\begin{figure}[h]
\includegraphics[width=\columnwidth]{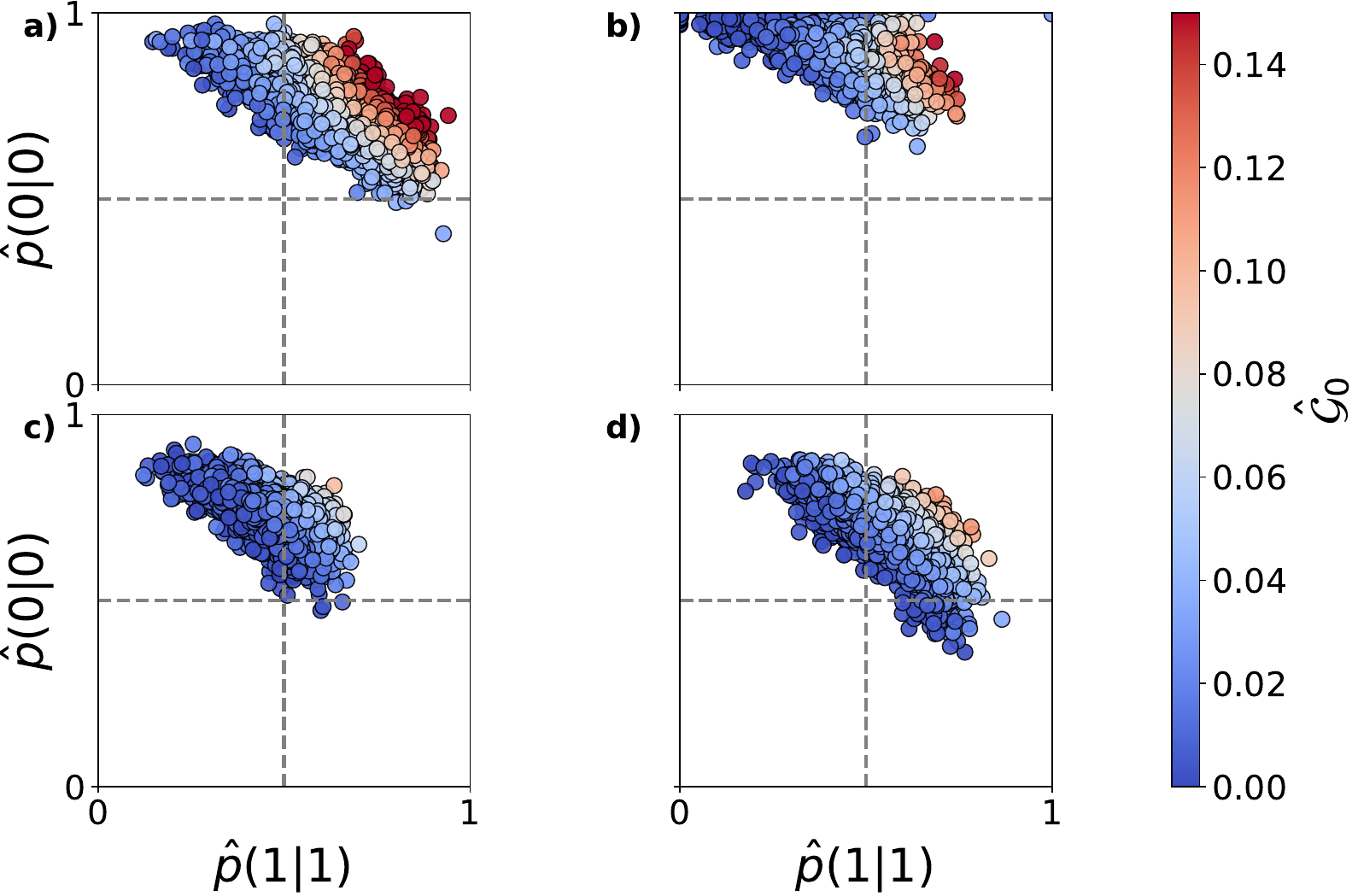}
 \caption{Estimated first-order transition probabilities for stations in the West Coast (top panels) and the Southeast (bottom panels). Winter months are shown in panels (a) and (c), and summer months in panels (b) and (d). Colors represent the corresponding value of $\hat{\mathcal{G}}_0$, with the colorbar saturated at $0.15$.}
 \label{fig:transprobs_seasons}
\end{figure}

In the top panels of Fig.~\ref{fig:transprobs_seasons}, we show the estimated transition probabilities, $\hat{p}(0|0)$ and $\hat{p}(1|1)$, for West Coast stations, with panel (a) corresponding to the winter months and panel (b) to the summer months. Colors indicate the corresponding value of $\hat{\mathcal{G}}_0$, with the colorbar saturated at $0.15$.
It can be observed that the stronger correlations in winter compared to summer in this region are consistent with overall higher values of $\hat{p}(1|1)$ and lower values of $\hat{p}(0|0)$.
These seasonal differences in transition probabilities are confirmed by Mann-Whitney U tests (p-values $<0.001$).

The higher winter tendency on the West Coast for a rainy day to be followed by another rainy day and dry days to be followed by a rainy day is consistent with the passage of frontal systems and atmospheric rivers that produce several consecutive wet periods~\cite{AtmosphericRiversasDroughtBustersontheUSWestCoast,FloodinginWesternWashingtonTheConnectiontoAtmosphericRivers}.

Similarly, the bottom panels of Fig.~\ref{fig:transprobs_seasons} show that in the Southeast the higher correlations in summer (panel (d)) compared to winter (panel (c)), already evident in Fig.~\ref{fig:seasons}, arise from both a greater persistence of rainy days (higher values of $\hat{p}(1|1)$) and an increased likelihood of a dry day being followed by a wet one (lower values of $\hat{p}(0|0)$) during the summer months.
Again, these seasonal differences in transition probabilities are confirmed by Mann-Whitney U tests (p-values $<0.001$).

This pattern is consistent with the Southeast summer wet-season regime, in which persistent subtropical circulation favors near-daily convective storms over extended periods~\cite{ChangestotheNorthAtlanticSubtropicalHighandItsRoleintheIntensificationofSummerRainfallVariabilityintheSoutheasternUnitedStates}.

In both the West Coast and the Southeast, the proposed method highlights that predictability arises from the persistence of weather regimes that tends to cluster precipitation events, underscoring its ability to identify wet-season patterns consistent with established climatological understanding.

\section{Conclusions}
\label{sec:conclusions}

In this work, we introduced an information-theoretic methodology to estimate the memory of stochastic processes based on the concept of predictability gain, defined as the negative discrete second derivative of block entropy. This quantity was shown to satisfy key properties that make it ideal to provide not only a rigorous criterion for determining memory order but also an interpretable measure of short-term temporal correlations. It should be noted that these methodological results are valid for discrete random variables. Extensions to continuous processes using differential entropy are not straightforward~\cite{cover} and constitute an interesting direction for future work. 

In particular, we proposed the Predictabilty Gain (PG) method to estimate the memory value of a process given a sequence of observations that combines bootstrap resampling and hypothesis testing. This allows us to
decide whether the predictability gain of the original sequence is consistent with what
would be expected if the sequence had memory $\eta$.
Applying Fisher's method to compute a combined p-value, we determine the smallest value of $\eta$ that aligns with the data, which is then selected as the memory of the process.

Extensive simulations demonstrated that the PG estimator outperforms classical model-selection criteria such as AIC and BIC, as we observed that our estimator is generally more robust, with its
accuracy being highest in most cases considered and its performance rapidly improving for larger data samples. 

This new estimator is independent of model selection and, consequently, it allows for a clear and robust
interpretation of the results.
Additionally, the PG estimator takes into consideration the
possibility that the data being analyzed may not be compatible with any of the memory orders being tested. However, the effectiveness of the proposed estimator may depend
on the number of resampled sequences used for comparison, and the critical value
chosen for the combined p-value threshold.

The advantages of the proposed approach become clearer in direct applications such as precipitation occurrence. Previous studies have estimated the order of precipitation sequences using the BIC estimator~\cite{https://doi.org/10.1002/joc.7175,schoof2008proper}, yet our simulations show that the PG estimator provides more reliable results. Additionally, quantifying the predictability gain provides a measure of temporal correlations, helping to develop models that are both accurate and efficient. This can guide responsible approximations, allowing model complexity to be reduced without compromising predictive skill and decreasing computational costs.
Moreover, applied in specific regions and months, our methodology can help decide when and where these models and approximations are reliable.

Applied to daily precipitation sequences across the contiguous United States, the PG estimator revealed that Markov chains of order $0$ and $1$ are overwhelmingly predominant, with higher-order dependencies rarely detected. The analysis of $\hat{\mathcal{G}}_0$ across space and time showed clear regional and seasonal patterns. In winter, strongest correlations occur along the West Coast, consistent with the passage of frontal systems and atmospheric rivers that produce consecutive wet days. Conversely, in summer, peak correlations are exhibited in the Southeast, reflecting the influence of subtropical circulation patterns that drive near-daily convective storms. These findings align with established climatological understanding and highlight the method's ability to uncover the persistence of weather regimes from occurrence data alone.

Overall, the framework developed here provides a powerful and interpretable approach for detecting and quantifying temporal dependence in finite sequences. Beyond precipitation, it offers a general tool for studying correlations in complex systems, with potential applications in social interactions, ecology, neuroscience, climate risk assessment, artificial intelligence and other fields where understanding short-term dependencies is essential.

\begin{acknowledgments}
Partial financial support has been received from Grants PID2021-122256NB-C21/C22 and PID2024-157493NB-C21/C22 funded by MICIU/AEI/10.13039/501100011033 and by “ERDF/EU”, and the María de Maeztu Program for units of Excellence in R\&D, grant CEX2021-001164-M.
\end{acknowledgments}

\section*{AUTHOR DECLARATIONS}

The authors have no conflicts to disclose.

\section*{DATA AVAILABILITY}

The data that support the findings of this study are openly available in the Global Historical Climatology Network Daily Database, reference number~\cite{menne2012overview}.

\bibliography{references}

\newpage
\appendix
\onecolumngrid

\renewcommand\thetable{\thesection.\arabic{table}} 
\setcounter{table}{0} 
\renewcommand\theHtable{Appendix.\thetable}

\section{Proof of properties}
\label{sec:appA}

\subsection{Proof of Proposition~\ref{prop:additive}}

Setting $k=u+1$ in Eq.~\eqref{eq:cummulative_gain} and comparing it with Eq.~\eqref{eq:pred_gain2}, it is clear that $G(u\to u+1)=\mathcal{G}_u$. 
Additionally, given the expression of $D_{\text{KL}}$ in Eq.~\eqref{eq:kl}, we can write for $k>u$,
\begin{equation}
\begin{split}
G(u\to k) = \sum_{i_1,\ldots,i_{k+1}=0}^{L-1}p(\beta_{i_1},\ldots,\beta_{i_{k+1}})
\ln \left(\dfrac{p(\beta_{i_{k+1}}|\beta_{i_1},\ldots,\beta_{i_{k}})}{p(\beta_{i_{k+1}}|\beta_{i_{k-u+1}},\ldots,\beta_{i_{k}})}\right),
\end{split}
\label{eq:cummulative_gain2}
\end{equation}

In general,
\begin{equation}
\begin{split}
\dfrac{p(\beta_{i_{k+1}}|\beta_{i_1},\ldots,\beta_{i_{k}})}{p(\beta_{i_{k+1}}|\beta_{i_{k-u+1}},\ldots,\beta_{i_{k}})} = 
\dfrac{p(\beta_{i_{k+1}}|\beta_{i_1},\ldots,\beta_{i_{k}})}{p(\beta_{i_{k+1}}|\beta_{i_{k-u}},\ldots,\beta_{i_{k}})}\dfrac{p(\beta_{i_{k+1}}|\beta_{i_{k-u}},\ldots,\beta_{i_{k}})}{p(\beta_{i_{k+1}}|\beta_{i_{k-u+1}},\ldots,\beta_{i_{k}})}.
\end{split}
\end{equation}
Plugging this into Eq.~\eqref{eq:cummulative_gain2} we get
\begin{equation}
\begin{split}
G(u\to k) &= \sum_{i_1,\ldots,i_{k+1}=0}^{L-1}p(\beta_{i_1},\ldots,\beta_{i_{k+1}}) 
\ln \left(\dfrac{p(\beta_{i_{k+1}}|\beta_{i_1},\ldots,\beta_{i_{k}})}{p(\beta_{i_{k+1}}|\beta_{i_{k-u}},\ldots,\beta_{i_{k}})}\right) \\
&+\sum_{i_1,\ldots,i_{k+1}=0}^{L-1}p(\beta_{i_1},\ldots,\beta_{i_{k+1}}) 
\ln \left(\dfrac{p(\beta_{i_{k+1}}|\beta_{i_{k-u}},\ldots,\beta_{i_{k}})}{p(\beta_{i_{k+1}}|\beta_{i_{k-u+1}},\ldots,\beta_{i_{k}})}\right).
\end{split}
\label{eq:cummulative_gain3}
\end{equation}
The first sum in Eq.~\eqref{eq:cummulative_gain3} is equal to $G(u+1\to k)$, whereas the second one, after summing $p(\beta_{i_1},\ldots,\beta_{i_{k+1}})$ over $i_1,\ldots,i_{k-u-1}$ and shifting the indices by $k-u$, is equal to $\mathcal{G}_u$. Thus,
\begin{equation}
G(u\to k) = G(u+1\to k) + \mathcal{G}_u.
\label{eq:cummulative_gain4}
\end{equation}

We can apply the same procedure over and over again until we reach
\begin{equation}
\begin{split}
G(u\to k) = \mathcal{G}_u + \mathcal{G}_{u+1}+\ldots+\mathcal{G}_{k-2}+G(k-1\to k) 
= \sum_{l=u}^{k-1}\mathcal{G}_l,
\end{split}
\end{equation}
which is the result we wanted to prove.

\subsection{Proof of Proposition~\ref{prop:total_pred_gain}}

Note that $G_T$, defined in Eq.~\eqref{eq:total_gain}, can be written as
\begin{equation}
\begin{split}
G_T &= \lim_{k\to \infty} G(0\to k) \\
&= \lim_{k\to \infty} \sum_{i_1,\ldots,i_{k+1}=0}^{L-1}p(\beta_{i_1},\ldots,\beta_{i_{k+1}}) \ln \left(\dfrac{p(\beta_{i_{k+1}}|\beta_{i_1},\ldots,\beta_{i_{k}})}{p(\beta_{i_{k+1}})}\right) \\
&= \lim_{k\to \infty} \left(\sum_{i_1,\ldots,i_{k+1}=0}^{L-1}p(\beta_{i_1},\ldots,\beta_{i_{k+1}}) \ln (p(\beta_{i_{k+1}}|\beta_{i_1},\ldots,\beta_{i_{k}})) -
\sum_{i_{k+1}=0}^{L-1}p(\beta_{i_{k+1}})\ln(p(\beta_{i_{k+1}}))\right).
\end{split}
\end{equation}

It can be shown~\cite{cover} that the entropy rate of a stationary process can be expressed as
\begin{equation}
 h = \lim_{k\to \infty} -\sum_{i_1,\ldots,i_{k+1}=0}^{L-1}p(\beta_{i_1},\ldots,\beta_{i_{k+1}}) \ln (p(\beta_{i_{k+1}}|\beta_{i_1},\ldots,\beta_{i_{k}})).
\end{equation}
Hence, we obtain the desired result
\begin{equation}
 G_T = H_1-h.
\end{equation}

\subsection{Proof of Proposition~\ref{prop:bounded}}

Given that the Kullback-Leibler divergence is always positive, it follows from Eq.~\eqref{eq:pred_gain2} that $\mathcal{G}_u\geq 0$, for $u\geq 0$. 

From Eq.~\eqref{eq:G_T}, since $h\geq 0$, we get that
\begin{equation}
\sum_{u=0}^{\infty} \mathcal{G}_u \leq H_1 \leq \ln(L).
\label{eq:bound}
\end{equation}
We just stated that $\mathcal{G}_u$ is always positive. Thus, Eq.~\eqref{eq:bound} can only hold if $\mathcal{G}_u \leq \ln(L)$ for all $u\geq 0$.

\subsection{Proof of Proposition~\ref{prop:pred_gain_distance}}

As previously stated, for a process with memory $m$, $H_r$ is linear for $r\geq m$. Thus, we can write
\begin{equation}
H_r = ar+b, \quad r\geq m.
\label{eq:line}
\end{equation}
We can extend this line to all values of $r$ as
\begin{equation}
\mathcal{H}(r) = ar+b,
\end{equation}
which, by definition, fulfills that $\mathcal{H}(r) = H_r$ if $r\geq m$.

Using Eq.~\eqref{eq:pred_gain} we can write
\begin{equation}
\mathcal{G}_{m-1} = -H_{m+1}+2H_m-H_{m-1}.
\label{eq:G_m}
\end{equation}
Replacing the values of $H_{m+1}$ and $H_m$ in Eq.~\eqref{eq:G_m} by their corresponding values given by Eq.~\eqref{eq:line} we find
\begin{equation}
\begin{split}
\mathcal{G}_{m-1} &= -a(m+1)-b + 2am + 2b - H_{m-1} \\
&= a(m-1)+b-H_{m-1} \\
&= \mathcal{H}(m-1)-H_{m-1},
\end{split}
\end{equation}
which corresponds to the Euclidean distance between the curves $\mathcal{H}(r)$ and $H_r$ at $r=m-1$.

\section{Zeros of the predictability gain}
\label{sec:AppB}

Given the $m$th-order transition probabilities of a Markov process of memory $m$, it is possible to calculate the probability of each block of size $m$ by solving the the following system of equations, obtained applying the law of total probability
\begin{equation}
p(\beta_{i_2},\ldots,\beta_{i_{m+1}}) = \sum_{i_1=0}^{L-1}p(\beta_{i_{m+1}}|\beta_{i_1},\ldots,\beta_{i_m})p(\beta_{i_1},\ldots,\beta_{i_m}),
\label{eq:system}
\end{equation}
for $i_2,\ldots,i_{m+1}=0,\ldots,L-1$. 

\begin{proposition}
\label{prop:general}

For a process with memory $m$ whose transition probabilities of order $m$ satisfy

\noindent
\begin{equation}
p(\beta_{i_{m+1}}|\beta_{i_1},\beta_{i_2},\ldots,\beta_{i_m}) = p(\beta_{i_{m+1}}|\beta_{i_1},\beta_{i_{1+\tau}},\beta_{i_{1+2\tau}},\ldots,\beta_{i_{1+(\frac{m}{\tau}-1)\tau}}),
\label{eq:condition_2}
\end{equation}
where $\tau$ is a factor (or divisor) of $m$, then, $\mathcal{G}_u=0$ for all $0\leq u \leq m-1$, except if $u=k\tau -1$, with $k=1,\ldots,\frac{m}{\tau}$. For $\mathcal{G}_{k\tau -1}$, we may or may not obtain a value of zero.
\end{proposition}

The condition imposed in Eq.~\eqref{eq:condition_2} implies that the transition probabilities of order $m$ exhibit periodic dependence on the outcomes with a period $\tau$, starting from the $m$th previous outcome up to the $\tau$th outcome.

Note that there are two special cases: $\tau=1$ and $\tau=m$. The first one corresponds to the common scenario, where the $m$th-order transition probabilities can depend on all previous outcomes. In this situation, we already know that, in principle, $\mathcal{G}_u > 0$ for $0\leq u \leq m-1$.

The case $\tau = m$ corresponds to the one where the $m$th-order transition probabilities depend only on the $m$th previous outcome.

Note that if $m$ is a prime number, the previous two cases are the only possibilities.
We can now move to the case where $m$ has a factor $1 < \tau \leq m$.
Defining $K \equiv \frac{m}{\tau}-1$, and replacing the transition probabilities in Eq.~\eqref{eq:condition_2} into Eq.~\eqref{eq:system}, we get
\begin{equation}
p(\beta_{i_2},\ldots,\beta_{i_{m+1}}) = \sum_{i_1=0}^{L-1} p(\beta_{i_{m+1}}|\beta_{i_1},\beta_{i_{1+\tau}},\ldots,\beta_{i_{1+K\tau}})p(\beta_{i_1},\ldots,\beta_{i_{m}}).
\label{eq:general}
\end{equation}
Summing Eq.~\eqref{eq:general} over $i_2,\ldots,i_{\tau},i_{2+\tau},\ldots,i_{m}$, we arrive at
\begin{equation}
p(\beta_{i_{1+\tau}},\beta_{i_{1+2\tau}},\ldots,\beta_{i_{m+1}}) = \sum_{i_1=0}^{L-1} p(\beta_{i_{m+1}}|\beta_{i_1},\beta_{i_{1+\tau}},\ldots,\beta_{i_{1+K\tau}})p(\beta_{i_1},\beta_{i_{1+\tau}},\ldots,\beta_{i_{1+K\tau}}).
\label{eq:general2}
\end{equation}
Observe that $m=\frac{m}{\tau}\tau = (K+1)\tau$. Thus, the probabilities on both sides of Eq.~\eqref{eq:general2} depend on $m/\tau$ outcomes, each separated by a step $\tau$.

Solving Eq.~\eqref{eq:general2} is equivalent to finding the steady states of a Markov process of order $m/\tau$. Therefore, we know we can find unique solutions to this equation~\cite{grinstead2012introduction}.

We will now show that
\begin{equation}
p(\beta_{i_1},\ldots,\beta_{i_{m}}) = \prod_{k=1}^{\tau} p(\beta_{i_k},\beta_{i_{k+\tau}},\ldots,\beta_{i_{k+K\tau}}),
\label{eq:general_sol}
\end{equation}
solves Eq.~\eqref{eq:general}.

Plugging Eq.~\eqref{eq:general_sol} into the right-hand side of Eq.~\eqref{eq:general}, we have

\noindent
\begin{equation}
\begin{split}
&\sum_{i_1=0}^{L-1} p(\beta_{i_{m+1}}|\beta_{i_1},\beta_{i_{1+\tau}},\ldots,\beta_{i_{1+K\tau}})\prod_{k=1}^{\tau} p(\beta_{i_k},\beta_{i_{k+\tau}},\ldots,\beta_{i_{k+K\tau}}) \\
&= \prod_{k=2}^{\tau} p(\beta_{i_k},\beta_{i_{k+\tau}},\ldots,\beta_{i_{k+K\tau}}) \sum_{i_1=0}^{L-1} p(\beta_{i_{m+1}}|\beta_{i_1},\beta_{i_{1+\tau}},\ldots,\beta_{i_{1+K\tau}})p(\beta_{i_1},\beta_{i_{1+\tau}},\ldots,\beta_{i_{1+K\tau}}) \\
&= \prod_{k=2}^{\tau} p(\beta_{i_k},\beta_{i_{k+\tau}},\ldots,\beta_{i_{k+K\tau}})p(\beta_{i_{1+\tau}},\beta_{i_{1+2\tau}},\ldots,\beta_{i_{m+1}}) \\
&= \prod_{k=2}^{\tau} p(\beta_{i_k},\beta_{i_{k+\tau}},\ldots,\beta_{i_{k+K\tau}})p(\beta_{i_{1+\tau}},\beta_{i_{1+2\tau}},\ldots,\beta_{i_{1+(K+1)\tau}}) \\
&= \prod_{k=2}^{\tau+1} p(\beta_{i_k},\beta_{i_{k+\tau}},\ldots,\beta_{i_{k+K\tau}})
\end{split}
\label{eq:general_sol2}
\end{equation}

Shifting all indices in Eq.~\eqref{eq:general_sol} by $1$, the left-hand side of Eq.~\eqref{eq:general} can be written as
\begin{equation}
\begin{split}
p(\beta_{i_2},\ldots,\beta_{i_{m+1}}) = \prod_{k=2}^{\tau+1} p(\beta_{i_k},\beta_{i_{k+\tau}},\ldots,\beta_{i_{k+K\tau}}),
\end{split}
\label{eq:general_sol3}
\end{equation}
Combining Eqs.~\eqref{eq:general_sol2} and~\eqref{eq:general_sol3} we observe that Eq.~\eqref{eq:general_sol} indeed solves Eq.~\eqref{eq:general}. 

We can observe in Eq.~\eqref{eq:general_sol} that the values $\beta_{i_1}$ and $\beta_{i_2}$ are in different factors. Therefore, $p(\beta_{i_1},\beta_{i_2})=p(\beta_{i_1})p(\beta_{i_2})$, which implies $\mathcal{G}_u=0$.
Moreover, if $1\leq u \leq \tau-1$, then the $u$th-order transition probabilities take the form
\begin{equation}
p(\beta_{i_{u+1}}|\beta_{i_1},\ldots,\beta_{i_u}) = \dfrac{\prod_{k=1}^{u+1}p(\beta_{i_k})}{\prod_{k=1}^{u}p(\beta_{i_k})}=p(\beta_{i_{u+1}}),
\end{equation}
which do not depend on the value of $\beta_{i_1}$. Therefore $\mathcal{G}_u=0$ for $u=0,1,\ldots \tau-2$.
However, the transition probabilities of order $\tau$ can be written as
\begin{equation}
p(\beta_{i_{\tau+1}}|\beta_{i_1},\ldots,\beta_{i_{\tau}}) = \dfrac{p(\beta_{i_1},\ldots,\beta_{i_{\tau+1}})}{p(\beta_{i_1},\ldots,\beta_{i_{\tau}})} = \dfrac{p(\beta_{i_1},\beta_{i_{\tau+1}})}{p(\beta_{i_1})},
\end{equation}
which does depend on $\beta_{i_1}$. Consequently, in principle, $\mathcal{G}_{\tau-1}>0$.

Notice that if $\tau+1\leq u \leq 2\tau-1$, then 
\begin{equation}
p(\beta_{i_{u+1}}|\beta_{i_1},\ldots,\beta_{i_u}) = \dfrac{p(\beta_{i_1},\beta_{i_{\tau+1}})A}{p(\beta_{i_1},\beta_{i_{\tau+1}})B},
\end{equation}
where the explicit form of the factors $A$ and $B$ is not relevant, except for the fact that they do not depend on $\beta_{i_1}$. Therefore $\mathcal{G}_u=0$ for $u=\tau,\tau+1,\ldots 2\tau-2$. However,
\begin{equation}
p(\beta_{i_{2\tau+1}}|\beta_{i_1},\ldots,\beta_{i_{2\tau}}) = \dfrac{p(\beta_{i_1},\beta_{i_{\tau+1}},\beta_{i_{2\tau+1}})}{p(\beta_{i_1},\beta_{i_{\tau+1}})},
\end{equation}
does depend on the value of $\beta_{i_1}$. Thus, $\mathcal{G}_{2\tau-1}>0$

Following this procedure, we can observe that $\mathcal{G}_u=0$, except if $u=k\tau -1$, with $k=1,\ldots,\frac{m}{\tau}$.

\begin{figure}[h]
 \centering

 \begin{minipage}{\linewidth}
 \begin{minipage}{0.05\linewidth}
  \raggedright \textbf{(a)}
 \end{minipage}
 \begin{minipage}{0.5\linewidth}
  \includegraphics[width=\linewidth]{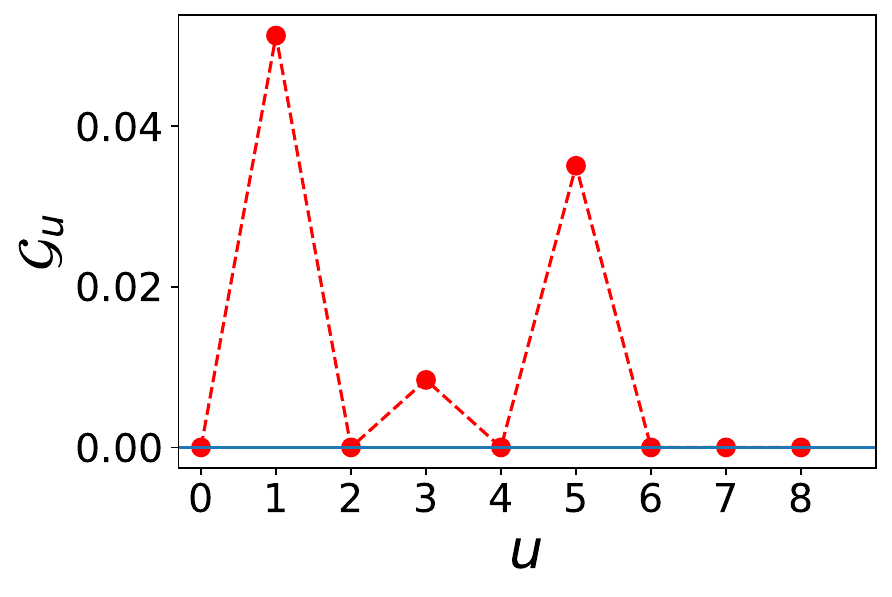}
 \end{minipage}
 \end{minipage}

 \vspace{2ex}

 \begin{minipage}{\linewidth}
 \begin{minipage}{0.05\linewidth}
  \raggedright \textbf{(b)}
 \end{minipage}
 \begin{minipage}{0.5\linewidth}
  \includegraphics[width=\linewidth]{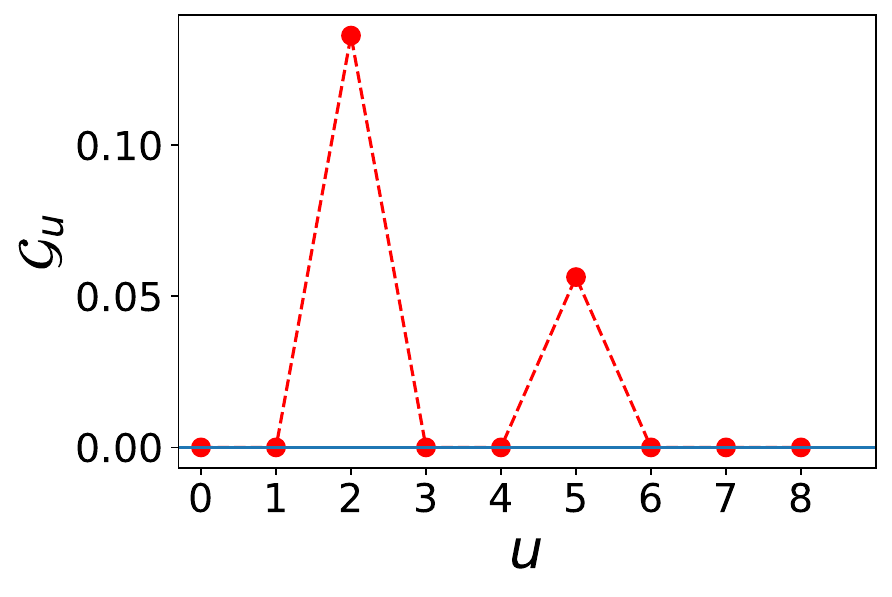}
 \end{minipage}
 \end{minipage}

 \caption{Predictability gain for two binary processes with memory $6$ whose transition probabilities satisfy Eq.~\eqref{eq:condition_2} with $\tau = 2$ in panel (a) and $\tau = 3$ in panel (b).}
 \label{fig:tau}
\end{figure}

We show in Fig.~\ref{fig:tau} two cases of binary systems with memory $m=6$ whose transition probabilities obey Eq.~\eqref{eq:condition_2} with $\tau = 2$ in panel (a) and $\tau = 3$ in panel (b). 
In the first case we observe that $\mathcal{G}_u=0$ for $u=0,2,4$, whereas in panel (b), $\mathcal{G}_u=0$ for $u=0,1,3,4$, as predicted.

As suggested by Proposition~\ref{prop:general}, processes that obey Eq.~\eqref{eq:condition_2} evolve in time as $\tau$ independent Markov chains of order $m/\tau$. These type of systems have been employed in recent applications~\cite{doi:10.1073/pnas.2105230118,mardt2022deep}.
However, these are not the only systems capable of producing zeros in the predictability gain.
In fact, identifying all instances where this occurs, for general values of $m$ and $L$, can be a very difficult task.

In the following proposition we present another case where we can observe this effect, which generalizes Proposition~\ref{prop:general} for $m=2$ and $L=2$.

\begin{proposition}

For a binary process ($\beta_0=0$ and $\beta_1=1$) with memory $m=2$, $\mathcal{G}_0=0$ if and only if the second-order transition probabilities obey
\begin{equation}
\dfrac{p(0|1,1)}{p(0|1,0)}=\dfrac{1-p(0|0,1)}{1-p(0|0,0)}.
\label{eq:condition_trans}
\end{equation}
\end{proposition}

First, observe that if $p(0|1,1)=p(0|1,0)$ and $p(0|0,1)=p(0|0,0)$, which clearly satisfy Eq.~\eqref{eq:condition_trans}, we have a system described by Eq.~\eqref{eq:condition_2} with $m=\tau=2$. We already showed that $\mathcal{G}_0=0$ in such case.

We will start by assuming that $\mathcal{G}_0=0$. This implies that $p(\beta_{i_1},\beta_{i_2})=p(\beta_{i_1})p(\beta_{i_2})$ for $i_1,i_2=0,1$. Plugging this into Eq.~\eqref{eq:system}, we have
\begin{equation}
p(\beta_{i_2},\beta_{i_3})=p(\beta_{i_2})p(\beta_{i_3})=\sum_{i_1=0}^1 p(\beta_{i_3}|\beta_{i_1},\beta_{i_2})p(\beta_{i_1})p(\beta_{i_2}).
\end{equation}
Thus,

\noindent
\begin{equation}
p(\beta_{i_3}) = \sum_{i_1=0}^1 p(\beta_{i_3}|\beta_{i_1},\beta_{i_2})p(\beta_{i_1}),
\label{eq:beta_3}
\end{equation}
for $i_2,i_3=0,1$. Setting $\beta_{i_3}=0$ in Eq.~\eqref{eq:beta_3} we get the following two equations, corresponding to $\beta_{i_2}=0,1$, respectively:
\begin{equation}
\begin{split}
p(0) &= p(0|0,0)p(0)+p(0|1,0)p(1), \\
p(0) &= p(0|0,1)p(0)+p(0|1,1)p(1).
\end{split}
\end{equation}
Replacing $p(1)=1-p(0)$, we get the following two expressions for $p(0)$:
\begin{equation}
\begin{split}
p(0) &= \dfrac{p(0|1,0)}{1-p(0|0,0)+p(0|1,0)}, \\
p(0) &= \dfrac{p(0|1,1)}{1-p(0|0,1)+p(0|1,1)}.
\label{eq:p(0)}
\end{split}
\end{equation}
Equating both formulas for $p(0)$ in Eq.~\eqref{eq:p(0)} and after some simple arithmetic, we find that the transition probabilities must satisfy Eq.~\eqref{eq:condition_trans}.
If we set $\beta_{i_3}=1$ in Eq.~\eqref{eq:beta_3} we arrive to the same condition.

The reverse result can be proven by showing that if the condition imposed by Eq.~\eqref{eq:condition_trans} is met, then $p(\beta_i,\beta_j)=p(\beta_i)p(\beta_j)$ for $\beta_i,\beta_j=0,1$, where $p(0)$ is given by Eq.~\eqref{eq:p(0)} and $p(1)=1-p(0)$, satisfy Eq.~\eqref{eq:system}. This implies that $\mathcal{G}_0=0$. 

\begin{figure}[h]
\centering
\includegraphics[width=.5\textwidth]{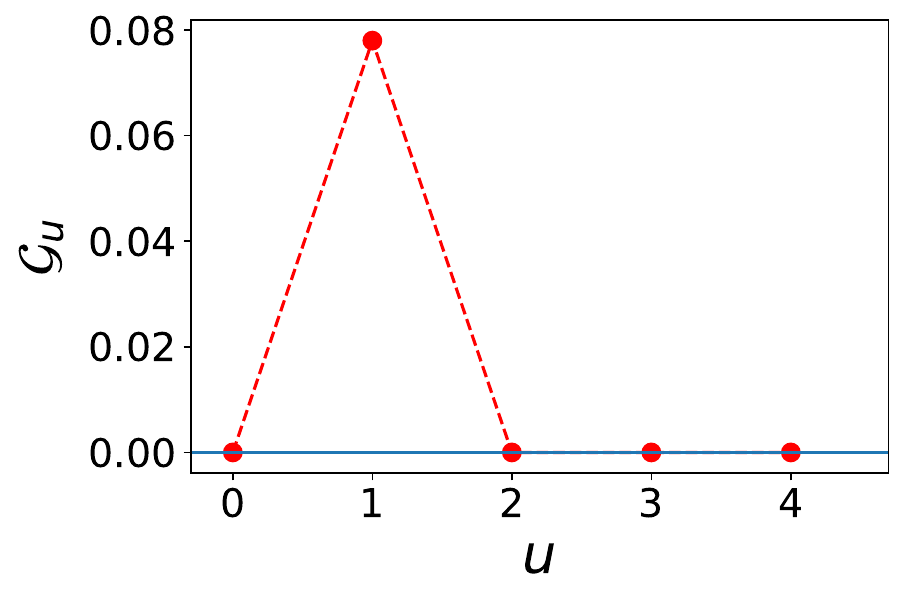}
 \caption{Predictability gain for a binary system with memory $m=2$ whose transition probabilities are $p(0|0,0)=0.5$, $p(0|0,1)=0.8$, $p(0|1,0)=0.6$ and $p(0|1,1)=0.24$, which satisfy Eq.~\eqref{eq:condition_trans}.}
 \label{fig:m=2}
\end{figure}

We show in Fig.~\ref{fig:m=2} a case where the transition probabilities are $p(0|0,0)=0.5$, $p(0|0,1)=0.8$, $p(0|1,0)=0.6$ and $p(0|1,1)=0.24$, which clearly satisfy Eq.~\eqref{eq:condition_trans}. We observe that indeed $\mathcal{G}_0=0$. 

\section{Fisher's combined p-value}
\label{app:fisher}

Fisher's method for combining p-values to test a global null hypothesis is based on the fact that, under the null hypothesis, p-values are uniformly distributed on the interval $[0,1]$ when the test statistic is continuous~\cite{doi:10.1080/00031305.1999.10474484}.
Furthermore, if $Q$ is a random variable uniformly distributed on $[0,1]$, then the variable $-2\ln(Q)$ follows a chi-squared distribution with $2$ degrees of freedom.
More generally, if $Y_1,\ldots,Y_M$ are independent chi-squared random variables with degrees of freedom $j_1,\ldots,j_M$, respectively, then the sum $Y_1+\ldots+Y_M$ follows a chi-squared distribution with $j_1+\ldots+j_M$ degrees of freedom~\cite{abramowitz1948handbook}.

From these results, it follows that when testing a global null hypothesis $\mathbf{N}$ composed of $M$ independent null hypotheses, whose corresponding p-values follow distributions $Q_1,\ldots,Q_M$, the combined test statistic
\begin{equation}
Y = -2\sum_{i=1}^M \ln(Q_i),
\end{equation}
follows a chi-squared distribution with $2M$ degrees of freedom when $\mathbf{N}$ is true.

Therefore, given a set of observed p-values $q_1,\ldots,q_M$, we compute the statistic
\begin{equation}
y = -2\sum_{i=1}^M \ln(q_i) = -2\ln(z),
\end{equation}
where $z=q_1\ldots q_M$. Fisher’s combined p-value can then be expressed as
\begin{equation}
q=P(Y\geq y \,|\, \mathbf{N}) = 1-I_M(-\ln(z)),
\label{eq:q1}
\end{equation}
with 
\begin{equation}
I_M(x) = \int_{0}^{2x}f_{2M}(t)dt, \quad x >0,
\label{eq:I_M}
\end{equation}
where $f_{2M}$ is the probability density function of a chi-squared distribution with $2M$ degrees of freedom, given by~\cite{abramowitz1948handbook}
\begin{equation}
f_{2M}(t) = \dfrac{t^{M-1} e^{-t/2}}{2^M(M-1)!}, \quad t\geq 0.
\end{equation}
Making the substitution $t\longrightarrow 2t$ in Eq.~\eqref{eq:I_M}, we get
\begin{equation}
I_M(x) = \dfrac{1}{(M-1)!}\int_0^x t^{M-1}e^{-t}dt.
\label{eq:IM2}
\end{equation}
From the equation above, it is straightforward to observe that 
\begin{equation}
I_1(x) = 1-e^{-x}.
\end{equation}

Moreover, integrating by parts Eq.~\eqref{eq:IM2} for $M\geq 2$, we see that
\begin{equation}
\begin{split}
I_M(x) &= -\dfrac{1}{(M-1)!}t^{M-1}e^{-t}\Big|_0^x + \dfrac{M-1}{(M-1)!}\int_0^x t^{M-2}e^{-t}dt \\
&= -\dfrac{x^{M-1}}{(M-1)!}e^{-x}+I_{M-1}(x).
\end{split}
\label{eq:IM3}
\end{equation}
Following this procedure in the right-hand side of Eq.~\eqref{eq:IM3} until reaching $I_1(x)$, we get
\begin{equation}
\begin{split}
I_M(x) &= -\dfrac{x^{M-1}}{(M-1)!}e^{-x} -\dfrac{x^{M-2}}{(M-2)!}e^{-x}-\ldots-xe^{-x}+I_1(x) \\
&= 1-e^{-x}\sum_{j=0}^{M-1}\dfrac{x^j}{j!}.
\end{split}
\label{eq:IM4}
\end{equation}
Finally, plugging Eq.~\eqref{eq:IM4} with $x=-\ln(z)$ in Eq.~\eqref{eq:q1}, we arrive at
\begin{equation}
q = z\sum_{j=0}^{M-1}\dfrac{(-\ln(z))^j}{j!}.
\end{equation}

\newpage
\section{Spatial distribution of estimated Markov memory per month}
\label{app:C}

\begin{figure}[h]
\includegraphics[width=\columnwidth]{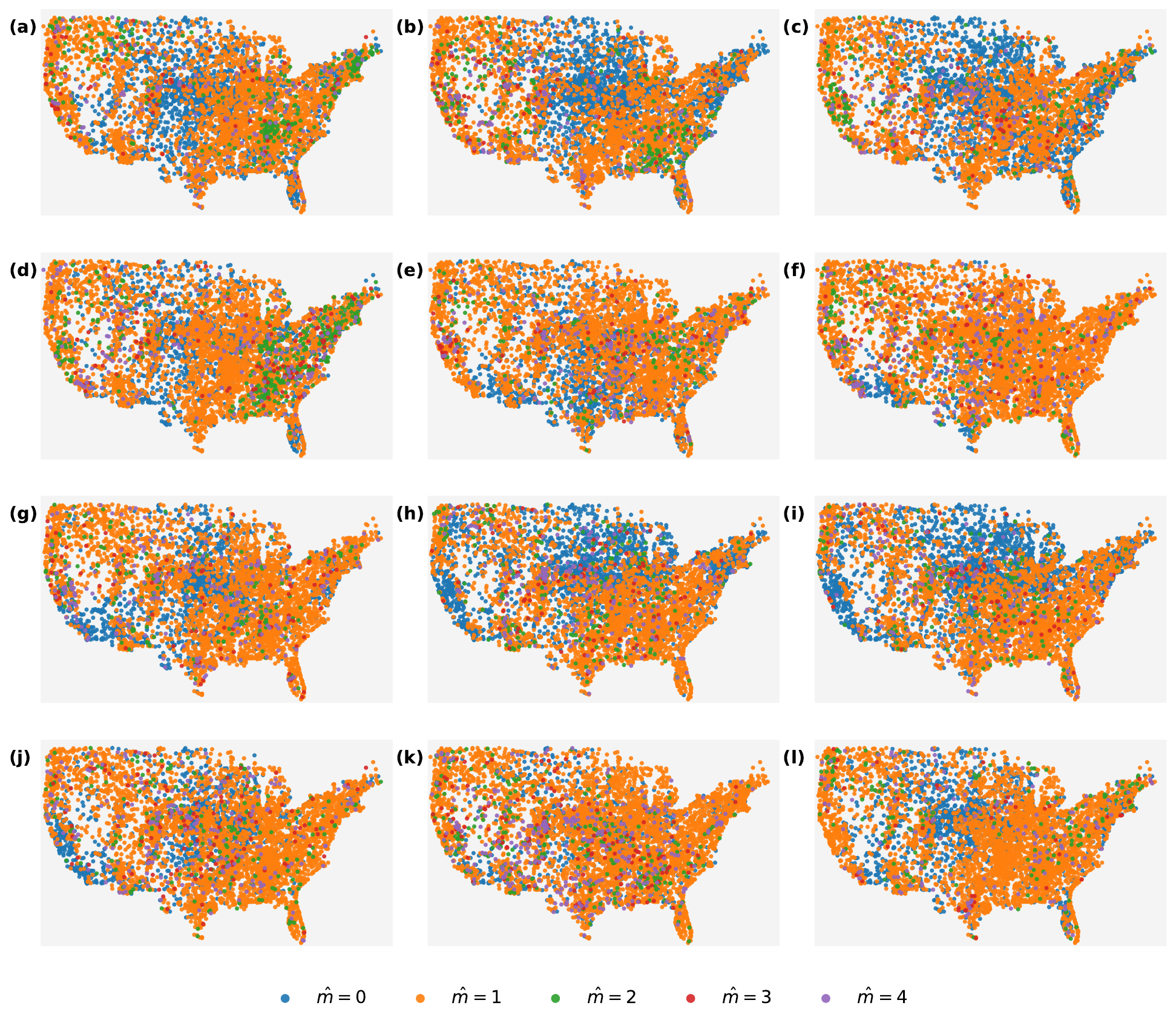}
 \caption{Spatial distribution of estimated memory $\hat{m}$ for precipitation occurrence across the contiguous United States. Each panel corresponds to a calendar month, with stations represented as dots colored according to their assigned memory order ($0-4$). Panel indices denote the corresponding month: (a) December, (b) January, (c) February, (d) March, (e) April, (f) May, (g) June, (h) July, (i) August, (j) September, (k) October, and (l) November.}
 \label{fig:spatial_month}
\end{figure}

\end{document}